\ifx\newheadisloaded\relax\immediate\write16{***already loaded}\endinput\else\let\newheadisloaded=\relax\fi
\gdef\PSfonts{T}
\magnification\magstep1

\newdimen\papwidth
\newdimen\papheight
\newskip\beforesectionskipamount  
\newskip\sectionskipamount 
\def\sectionskip{\vskip\sectionskipamount}
\def\beforesectionskip{\vskip\beforesectionskipamount}
\papwidth=16truecm
\papheight=22truecm
\voffset=0.4truecm
\hoffset=0.4truecm
\hsize=\papwidth
\vsize=\papheight
\nopagenumbers
\headline={\ifnum\pageno>1 {\hss\tenrm-\ \folio\ -\hss} \else
{\hfill}\fi}
\newdimen\texpscorrection
\texpscorrection=0.15truecm 

\def\sectionsize{\twelvepoint}
\def\sectiontype{\bf}
\def\subsectionsize{}
\def\subsectiontype{\bf}
\def\em{\sl}
\newfam\truecmsy
\newfam\truecmr
\newfam\msbfam
\newfam\scriptfam
\newfam\truecmsy
\newskip\ttglue 
\if F\PSfonts
\font\twelverm=cmr12
\font\tenrm=cmr10
\font\eightrm=cmr8
\font\sevenrm=cmr7
\font\sixrm=cmr6
\font\fiverm=cmr5

\font\twelvebf=cmbx12
\font\tenbf=cmbx10
\font\eightbf=cmbx8
\font\sevenbf=cmbx7
\font\sixbf=cmbx6
\font\fivebf=cmbx5

\font\twelveit=cmti12
\font\tenit=cmti10
\font\eightit=cmti8
\font\sevenit=cmti7
\font\sixit=cmti6
\font\fiveit=cmti5

\font\twelvesl=cmsl12
\font\tensl=cmsl10
\font\eightsl=cmsl8
\font\sevensl=cmsl7
\font\sixsl=cmsl6
\font\fivesl=cmsl5

\font\twelvei=cmmi12
\font\teni=cmmi10
\font\eighti=cmmi8
\font\seveni=cmmi7
\font\sixi=cmmi6
\font\fivei=cmmi5

\font\twelvesy=cmsy10	at	12pt
\font\tensy=cmsy10
\font\eightsy=cmsy8
\font\sevensy=cmsy7
\font\sixsy=cmsy6
\font\fivesy=cmsy5
\font\twelvetruecmsy=cmsy10	at	12pt
\font\tentruecmsy=cmsy10
\font\eighttruecmsy=cmsy8
\font\seventruecmsy=cmsy7
\font\sixtruecmsy=cmsy6
\font\fivetruecmsy=cmsy5

\font\twelvetruecmr=cmr12
\font\tentruecmr=cmr10
\font\eighttruecmr=cmr8
\font\seventruecmr=cmr7
\font\sixtruecmr=cmr6
\font\fivetruecmr=cmr5

\font\twelvebf=cmbx12
\font\tenbf=cmbx10
\font\eightbf=cmbx8
\font\sevenbf=cmbx7
\font\sixbf=cmbx6
\font\fivebf=cmbx5

\font\twelvett=cmtt12
\font\tentt=cmtt10
\font\eighttt=cmtt8

\font\twelveex=cmex10	at	12pt
\font\tenex=cmex10

\font\twelvemsb=msbm10	at	12pt
\font\tenmsb=msbm10
\font\eightmsb=msbm8
\font\sevenmsb=msbm7
\font\sixmsb=msbm6
\font\fivemsb=msbm5

\font\twelvescr=eusm10 at 12pt
\font\tenscr=eusm10
\font\eightscr=eusm8
\font\sevenscr=eusm7
\font\sixscr=eusm6
\font\fivescr=eusm5
\fi
\if T\PSfonts
\font\twelverm=ptmr	at	12pt
\font\tenrm=ptmr	at	10pt
\font\eightrm=ptmr	at	8pt
\font\sevenrm=ptmr	at	7pt
\font\sixrm=ptmr	at	6pt
\font\fiverm=ptmr	at	5pt

\font\twelvebf=ptmb	at	12pt
\font\tenbf=ptmb	at	10pt
\font\eightbf=ptmb	at	8pt
\font\sevenbf=ptmb	at	7pt
\font\sixbf=ptmb	at	6pt
\font\fivebf=ptmb	at	5pt

\font\twelveit=ptmri	at	12pt
\font\tenit=ptmri	at	10pt
\font\eightit=ptmri	at	8pt
\font\sevenit=ptmri	at	7pt
\font\sixit=ptmri	at	6pt
\font\fiveit=ptmri	at	5pt

\font\twelvesl=ptmro	at	12pt
\font\tensl=ptmro	at	10pt
\font\eightsl=ptmro	at	8pt
\font\sevensl=ptmro	at	7pt
\font\sixsl=ptmro	at	6pt
\font\fivesl=ptmro	at	5pt

\font\twelvei=cmmi12
\font\teni=cmmi10
\font\eighti=cmmi8
\font\seveni=cmmi7
\font\sixi=cmmi6
\font\fivei=cmmi5

\font\twelvesy=cmsy10	at	12pt
\font\tensy=cmsy10
\font\eightsy=cmsy8
\font\sevensy=cmsy7
\font\sixsy=cmsy6
\font\fivesy=cmsy5
\font\twelvetruecmsy=cmsy10	at	12pt
\font\tentruecmsy=cmsy10
\font\eighttruecmsy=cmsy8
\font\seventruecmsy=cmsy7
\font\sixtruecmsy=cmsy6
\font\fivetruecmsy=cmsy5

\font\twelvetruecmr=cmr12
\font\tentruecmr=cmr10
\font\eighttruecmr=cmr8
\font\seventruecmr=cmr7
\font\sixtruecmr=cmr6
\font\fivetruecmr=cmr5

\font\twelvebf=cmbx12
\font\tenbf=cmbx10
\font\eightbf=cmbx8
\font\sevenbf=cmbx7
\font\sixbf=cmbx6
\font\fivebf=cmbx5

\font\twelvett=cmtt12
\font\tentt=cmtt10
\font\eighttt=cmtt8

\font\twelveex=cmex10	at	12pt
\font\tenex=cmex10

\font\twelvemsb=msbm10	at	12pt
\font\tenmsb=msbm10
\font\eightmsb=msbm8
\font\sevenmsb=msbm7
\font\sixmsb=msbm6
\font\fivemsb=msbm5

\font\twelvescr=eusm10 at 12pt
\font\tenscr=eusm10
\font\eightscr=eusm8
\font\sevenscr=eusm7
\font\sixscr=eusm6
\font\fivescr=eusm5
\fi
\def\eightpoint{\def\rm{\fam0\eightrm}%
\textfont0=\eightrm
  \scriptfont0=\sixrm
  \scriptscriptfont0=\fiverm 
\textfont1=\eighti
  \scriptfont1=\sixi
  \scriptscriptfont1=\fivei 
\textfont2=\eightsy
  \scriptfont2=\sixsy
  \scriptscriptfont2=\fivesy 
\textfont3=\tenex
  \scriptfont3=\tenex
  \scriptscriptfont3=\tenex 
\textfont\itfam=\eightit
  \scriptfont\itfam=\sixit
  \scriptscriptfont\itfam=\fiveit 
  \def\it{\fam\itfam\eightit}%
\textfont\slfam=\eightsl
  \scriptfont\slfam=\sixsl
  \scriptscriptfont\slfam=\fivesl 
  \def\sl{\fam\slfam\eightsl}%
\textfont\ttfam=\eighttt
  \def\tt{\fam\ttfam\eighttt}%
\textfont\bffam=\eightbf
  \scriptfont\bffam=\sixbf
  \scriptscriptfont\bffam=\fivebf
  \def\bf{\fam\bffam\eightbf}%
\textfont\scriptfam=\eightscr
  \scriptfont\scriptfam=\sixscr
  \scriptscriptfont\scriptfam=\fivescr
  \def\script{\fam\scriptfam\eightscr}%
\textfont\msbfam=\eightmsb
  \scriptfont\msbfam=\sixmsb
  \scriptscriptfont\msbfam=\fivemsb
  \def\bb{\fam\msbfam\eightmsb}%
\textfont\truecmr=\eighttruecmr
  \scriptfont\truecmr=\sixtruecmr
  \scriptscriptfont\truecmr=\fivetruecmr
  \def\truerm{\fam\truecmr\eighttruecmr}%
\textfont\truecmsy=\eighttruecmsy
  \scriptfont\truecmsy=\sixtruecmsy
  \scriptscriptfont\truecmsy=\fivetruecmsy
\tt \ttglue=.5em plus.25em minus.15em 
\normalbaselineskip=9pt
\setbox\strutbox=\hbox{\vrule height7pt depth2pt width0pt}%
\normalbaselines
\rm
}

\def\tenpoint{\def\rm{\fam0\tenrm}%
\textfont0=\tenrm
  \scriptfont0=\sevenrm
  \scriptscriptfont0=\fiverm 
\textfont1=\teni
  \scriptfont1=\seveni
  \scriptscriptfont1=\fivei 
\textfont2=\tensy
  \scriptfont2=\sevensy
  \scriptscriptfont2=\fivesy 
\textfont3=\tenex
  \scriptfont3=\tenex
  \scriptscriptfont3=\tenex 
\textfont\itfam=\tenit
  \scriptfont\itfam=\sevenit
  \scriptscriptfont\itfam=\fiveit 
  \def\it{\fam\itfam\tenit}%
\textfont\slfam=\tensl
  \scriptfont\slfam=\sevensl
  \scriptscriptfont\slfam=\fivesl 
  \def\sl{\fam\slfam\tensl}%
\textfont\ttfam=\tentt
  \def\tt{\fam\ttfam\tentt}%
\textfont\bffam=\tenbf
  \scriptfont\bffam=\sevenbf
  \scriptscriptfont\bffam=\fivebf
  \def\bf{\fam\bffam\tenbf}%
\textfont\scriptfam=\tenscr
  \scriptfont\scriptfam=\sevenscr
  \scriptscriptfont\scriptfam=\fivescr
  \def\script{\fam\scriptfam\tenscr}%
\textfont\msbfam=\tenmsb
  \scriptfont\msbfam=\sevenmsb
  \scriptscriptfont\msbfam=\fivemsb
  \def\bb{\fam\msbfam\tenmsb}%
\textfont\truecmr=\tentruecmr
  \scriptfont\truecmr=\seventruecmr
  \scriptscriptfont\truecmr=\fivetruecmr
  \def\truerm{\fam\truecmr\tentruecmr}%
\textfont\truecmsy=\tentruecmsy
  \scriptfont\truecmsy=\seventruecmsy
  \scriptscriptfont\truecmsy=\fivetruecmsy
\tt \ttglue=.5em plus.25em minus.15em 
\normalbaselineskip=12pt
\setbox\strutbox=\hbox{\vrule height8.5pt depth3.5pt width0pt}%
\normalbaselines
\rm
}

\def\twelvepoint{\def\rm{\fam0\twelverm}%
\textfont0=\twelverm
  \scriptfont0=\tenrm
  \scriptscriptfont0=\eightrm 
\textfont1=\twelvei
  \scriptfont1=\teni
  \scriptscriptfont1=\eighti 
\textfont2=\twelvesy
  \scriptfont2=\tensy
  \scriptscriptfont2=\eightsy 
\textfont3=\twelveex
  \scriptfont3=\twelveex
  \scriptscriptfont3=\twelveex 
\textfont\itfam=\twelveit
  \scriptfont\itfam=\tenit
  \scriptscriptfont\itfam=\eightit 
  \def\it{\fam\itfam\twelveit}%
\textfont\slfam=\twelvesl
  \scriptfont\slfam=\tensl
  \scriptscriptfont\slfam=\eightsl 
  \def\sl{\fam\slfam\twelvesl}%
\textfont\ttfam=\twelvett
  \def\tt{\fam\ttfam\twelvett}%
\textfont\bffam=\twelvebf
  \scriptfont\bffam=\tenbf
  \scriptscriptfont\bffam=\eightbf
  \def\bf{\fam\bffam\twelvebf}%
\textfont\scriptfam=\twelvescr
  \scriptfont\scriptfam=\tenscr
  \scriptscriptfont\scriptfam=\eightscr
  \def\script{\fam\scriptfam\twelvescr}%
\textfont\msbfam=\twelvemsb
  \scriptfont\msbfam=\tenmsb
  \scriptscriptfont\msbfam=\eightmsb
  \def\bb{\fam\msbfam\twelvemsb}%
\textfont\truecmr=\twelvetruecmr
  \scriptfont\truecmr=\tentruecmr
  \scriptscriptfont\truecmr=\eighttruecmr
  \def\truerm{\fam\truecmr\twelvetruecmr}%
\textfont\truecmsy=\twelvetruecmsy
  \scriptfont\truecmsy=\tentruecmsy
  \scriptscriptfont\truecmsy=\eighttruecmsy
\tt \ttglue=.5em plus.25em minus.15em 
\setbox\strutbox=\hbox{\vrule height7pt depth2pt width0pt}%
\normalbaselineskip=15pt
\normalbaselines
\rm
}
%
\fontdimen16\tensy=2.7pt
\fontdimen13\tensy=4.3pt
\fontdimen17\tensy=2.7pt
\fontdimen14\tensy=4.3pt
\fontdimen18\tensy=4.3pt
\fontdimen16\eightsy=2.7pt
\fontdimen13\eightsy=4.3pt
\fontdimen17\eightsy=2.7pt
\fontdimen14\eightsy=4.3pt
\fontdimen18\sevensy=4.3pt
\fontdimen16\sevensy=1.8pt
\fontdimen13\sevensy=4.3pt
\fontdimen17\sevensy=2.7pt
\fontdimen14\sevensy=4.3pt
\fontdimen18\sevensy=4.3pt
%
\def\hexnumber#1{\ifcase#1 0\or1\or2\or3\or4\or5\or6\or7\or8\or9\or
 A\or B\or C\or D\or E\or F\fi}
\mathcode`\=="3\hexnumber\truecmr3D
\mathchardef\not="3\hexnumber\truecmsy36
\mathcode`\+="2\hexnumber\truecmr2B
\mathcode`\(="4\hexnumber\truecmr28
\mathcode`\)="5\hexnumber\truecmr29
\mathcode`\!="5\hexnumber\truecmr21
\mathcode`\(="4\hexnumber\truecmr28
\mathcode`\)="5\hexnumber\truecmr29

\def\hat{\mathaccent"0\hexnumber\truecmr5E }

\def\Phi{\mathchar"0\hexnumber\truecmr08 }
\def\Gamma {\mathchar"0\hexnumber\truecmr00 }
\def\Delta {\mathchar"0\hexnumber\truecmr01 }
\def\Theta {\mathchar"0\hexnumber\truecmr02 }
\def\Lambda{\mathchar"0\hexnumber\truecmr03 }
\def\Xi {\mathchar"0\hexnumber\truecmr04 }
\def\Pi{\mathchar"0\hexnumber\truecmr05 }
\def\Sigma{\mathchar"0\hexnumber\truecmr06 }
\def\Upsilon {\mathchar"0\hexnumber\truecmr07 }
\def\Phi {\mathchar"0\hexnumber\truecmr08 }
\def\Psi {\mathchar"0\hexnumber\truecmr09 }
\def\Omega{\mathchar"0\hexnumber\truecmr0A }
\newcount\EQNcount \EQNcount=1
\newcount\CLAIMcount \CLAIMcount=1
\newcount\SECTIONcount \SECTIONcount=0
\newcount\SUBSECTIONcount \SUBSECTIONcount=1
\def\ifff(#1,#2,#3){\ifundefined{#1#2}%
\expandafter\xdef\csname #1#2\endcsname{#3}\else%
\fi}
\def\NEWDEF #1,#2,#3 {\ifff({#1},{#2},{#3})}
\def\actualnumber{\number\SECTIONcount}
\def\EQ(#1){\lmargin(#1)\eqno\tag(#1)}
\def\NR(#1){&\lmargin(#1)\tag(#1)\cr}  
\def\tag(#1){\lmargin(#1)({\rm \actualnumber}.\number\EQNcount)
 \NEWDEF e,#1,(\actualnumber.\number\EQNcount)
\global\advance\EQNcount by 1
}
\def\SECT(#1)#2\par{\lmargin(#1)\SECTION#2\par
\NEWDEF s,#1,{\actualnumber}
}
\def\SUBSECT(#1)#2\par{\lmargin(#1)
\SUBSECTION#2\par 
\NEWDEF s,#1,{\actualnumber.\number\SUBSECTIONcount}
}
\def\CLAIM #1(#2) #3\par{
\vskip.1in\medbreak\noindent
{\lmargin(#2)\bf #1\ \actualnumber.\number\CLAIMcount.} {\sl #3}\par
\NEWDEF c,#2,{#1\ \actualnumber.\number\CLAIMcount}
\global\advance\CLAIMcount by 1
\ifdim\lastskip<\medskipamount
\removelastskip\penalty55\medskip\fi}
\def\CLAIMNONR #1(#2) #3\par{
\vskip.1in\medbreak\noindent
{\lmargin(#2)\bf #1.} {\sl #3}\par
\NEWDEF c,#2,{#1}
\global\advance\CLAIMcount by 1
\ifdim\lastskip<\medskipamount
\removelastskip\penalty55\medskip\fi}
\def\SECTION#1\par{\vskip0pt plus.3\vsize\penalty-75
    \vskip0pt plus -.3\vsize
    \global\advance\SECTIONcount by 1
    \beforesectionskip\noindent
{\sectionsize\sectiontype \actualnumber.\ #1}
    \EQNcount=1
    \CLAIMcount=1
    \SUBSECTIONcount=1
    \nobreak\sectionskip\noindent}
\def\SECTIONNONR#1\par{\vskip0pt plus.3\vsize\penalty-75
    \vskip0pt plus -.3\vsize
    \global\advance\SECTIONcount by 1
    \beforesectionskip\noindent
{\sectionsize\sectiontype  #1}
     \EQNcount=1
     \CLAIMcount=1
     \SUBSECTIONcount=1
     \nobreak\sectionskip\noindent}
\def\SUBSECTION#1\par{\vskip0pt plus.2\vsize\penalty-75%
    \vskip0pt plus -.2\vsize%
    \beforesectionskip\noindent%
{\subsectionsize\subsectiontype \actualnumber.\number\SUBSECTIONcount.\ #1}
    \global\advance\SUBSECTIONcount by 1
    \nobreak\sectionskip\noindent}
\def\SUBSECTIONNONR#1\par{\vskip0pt plus.2\vsize\penalty-75
    \vskip0pt plus -.2\vsize
\beforesectionskip\noindent
{\subsectionsize\subsectiontype #1}
    \nobreak\sectionskip\noindent\noindent}
\def\ifundefined#1{\expandafter\ifx\csname#1\endcsname\relax}
\def\equ(#1){\ifundefined{e#1}$\spadesuit$#1\else\csname e#1\endcsname\fi}
\def\clm(#1){\ifundefined{c#1}$\spadesuit$#1\else\csname c#1\endcsname\fi}
\def\sec(#1){\ifundefined{s#1}$\spadesuit$#1
\else Section \csname s#1\endcsname\fi}
\def\fig(#1){\ifundefined{fig#1}$\spadesuit$#1\else\csname fig#1\endcsname\fi}
\let\endarg=\par
\def\finish{\def\endarg{\par\endgroup}}
\def\start{\endarg\begingroup}

 \def\beginFROM{\start\parskip=0pt\vskip\baselineskip
\def\finish{\def\endarg{\egroup\par\endgroup}}
  \vbox\bgroup\obeylines\eightpoint\em\finish}

\def\ABSTRACT#1\par{
\vskip 1in {\noindent\sectionsize\sectiontype Abstract.} #1 \par}

\def\TODAY{\number\day~\ifcase\month\or January \or February \or March \or
April \or May \or June
\or July \or August \or September \or October \or November \or December \fi
\number\year\timecount=\number\time
\divide\timecount by 60
}
\newcount\timecount
\def\DRAFT{\def\lmargin(##1){\strut\vadjust{\kern-\strutdepth
\vtop to \strutdepth{
\baselineskip\strutdepth\vss\rlap{\kern-1.2 truecm\eightpoint{##1}}}}}
\font\footfont=cmti7
\footline={{\footfont \hfil File:\jobname, \TODAY,  \number\timecount h}}
}
\newbox\strutboxJPE
\setbox\strutboxJPE=\hbox{\strut}
\def\subitem#1#2\par{\vskip\baselineskip\vskip-\ht\strutboxJPE{\item{#1}#2}}
\gdef\strutdepth{\dp\strutbox}
\def\lmargin(#1){}
\def\period{\unskip.\spacefactor3000 { }}
%
%
\newbox\noboxJPE
\newbox\byboxJPE
\newbox\paperboxJPE
\newbox\yrboxJPE
\newbox\jourboxJPE
\newbox\pagesboxJPE
\newbox\volboxJPE
\newbox\preprintboxJPE
\newbox\toappearboxJPE
\newbox\bookboxJPE
\newbox\bybookboxJPE
\newbox\publisherboxJPE
\newbox\inprintboxJPE
\def\refclearJPE{
   \setbox\noboxJPE=\null             \gdef\isnoJPE{F}
   \setbox\byboxJPE=\null             \gdef\isbyJPE{F}
   \setbox\paperboxJPE=\null          \gdef\ispaperJPE{F}
   \setbox\yrboxJPE=\null             \gdef\isyrJPE{F}
   \setbox\jourboxJPE=\null           \gdef\isjourJPE{F}
   \setbox\pagesboxJPE=\null          \gdef\ispagesJPE{F}
   \setbox\volboxJPE=\null            \gdef\isvolJPE{F}
   \setbox\preprintboxJPE=\null       \gdef\ispreprintJPE{F}
   \setbox\toappearboxJPE=\null       \gdef\istoappearJPE{F}
   \setbox\inprintboxJPE=\null        \gdef\isinprintJPE{F}
   \setbox\bookboxJPE=\null           \gdef\isbookJPE{F}  \gdef\isinbookJPE{F}
     
   \setbox\bybookboxJPE=\null         \gdef\isbybookJPE{F}
   \setbox\publisherboxJPE=\null      \gdef\ispublisherJPE{F}
     
}
\def\widestlabel#1{\setbox0=\hbox{#1\enspace}\refindent=\wd0\relax}
\def\ref{\refclearJPE\bgroup}
\def\no   {\egroup\gdef\isnoJPE{T}\setbox\noboxJPE=\hbox\bgroup}
\def\by   {\egroup\gdef\isbyJPE{T}\setbox\byboxJPE=\hbox\bgroup}
\def\paper{\egroup\gdef\ispaperJPE{T}\setbox\paperboxJPE=\hbox\bgroup}
\def\yr{\egroup\gdef\isyrJPE{T}\setbox\yrboxJPE=\hbox\bgroup}
\def\jour{\egroup\gdef\isjourJPE{T}\setbox\jourboxJPE=\hbox\bgroup}
\def\pages{\egroup\gdef\ispagesJPE{T}\setbox\pagesboxJPE=\hbox\bgroup}
\def\vol{\egroup\gdef\isvolJPE{T}\setbox\volboxJPE=\hbox\bgroup\bf}
\def\preprint{\egroup\gdef
\ispreprintJPE{T}\setbox\preprintboxJPE=\hbox\bgroup}
\def\toappear{\egroup\gdef
\istoappearJPE{T}\setbox\toappearboxJPE=\hbox\bgroup}
\def\inprint{\egroup\gdef
\isinprintJPE{T}\setbox\inprintboxJPE=\hbox\bgroup}
\def\book{\egroup\gdef\isbookJPE{T}\setbox\bookboxJPE=\hbox\bgroup\em}
\def\publisher{\egroup\gdef
\ispublisherJPE{T}\setbox\publisherboxJPE=\hbox\bgroup}
\def\inbook{\egroup\gdef\isinbookJPE{T}\setbox\bookboxJPE=\hbox\bgroup\em}
\def\bybook{\egroup\gdef\isbybookJPE{T}\setbox\bybookboxJPE=\hbox\bgroup}
\newdimen\refindent
\refindent=5em
\def\endref{\egroup \sfcode`.=1000
 \if T\isnoJPE
 \hangindent\refindent\hangafter=1
      \noindent\hbox to\refindent{[\unhbox\noboxJPE\unskip]\hss}\ignorespaces
     \else  \noindent    \fi
 \if T\isbyJPE    \unhbox\byboxJPE\unskip: \fi
 \if T\ispaperJPE \unhbox\paperboxJPE\unskip\period \fi
 \if T\isbookJPE {\it\unhbox\bookboxJPE\unskip}\if T\ispublisherJPE, \else.
\fi\fi
 \if T\isinbookJPE In {\it\unhbox\bookboxJPE\unskip}\if T\isbybookJPE,
\else\period \fi\fi
 \if T\isbybookJPE  (\unhbox\bybookboxJPE\unskip)\period \fi
 \if T\ispublisherJPE \unhbox\publisherboxJPE\unskip \if T\isjourJPE, \else\if
T\isyrJPE \  \else\period \fi\fi\fi
 \if T\istoappearJPE (To appear)\period \fi
 \if T\ispreprintJPE Pre\-print\period \fi
 \if T\isjourJPE    \unhbox\jourboxJPE\unskip\ \fi
 \if T\isvolJPE     \unhbox\volboxJPE\unskip\if T\ispagesJPE, \else\ \fi\fi
 \if T\ispagesJPE   \unhbox\pagesboxJPE\unskip\  \fi
 \if T\isyrJPE      (\unhbox\yrboxJPE\unskip)\period \fi
 \if T\isinprintJPE (in print)\period \fi
\filbreak
}
\def\hexnumber#1{\ifcase#1 0\or1\or2\or3\or4\or5\or6\or7\or8\or9\or
 A\or B\or C\or D\or E\or F\fi}
\textfont\msbfam=\tenmsb
\scriptfont\msbfam=\sevenmsb
\scriptscriptfont\msbfam=\fivemsb
\mathchardef\varkappa="0\hexnumber\msbfam7B
\newcount\FIGUREcount \FIGUREcount=0
\newdimen\figcenter
\def\definefigure#1{\global\advance\FIGUREcount by 1%
\NEWDEF fig,#1,{Fig.\ \number\FIGUREcount}
\immediate\write16{  FIG \number\FIGUREcount : #1}}
\def\figurewithtexplus #1 #2 #3 #4 #5 #6\cr{\null%
\definefigure{#1}
{\goodbreak\figcenter=\hsize\relax
\advance\figcenter by -#4truecm
\divide\figcenter by 2
\midinsert\noindent
{\centerline{\vbox{\hrule height 0.6pt
	\hbox{\vrule width 0.6pt height #3truecm \kern 14truecm
		\vrule width 0.6pt}
	\hrule height 0.6pt}}}\vskip 0.8truecm\par\noindent 
\vbox{\eightpoint\noindent
{\bf\fig(#1)}: #6}\endinsert}}
\catcode`@=11
\def\footnote#1{\let\@sf\empty 
  \ifhmode\edef\@sf{\spacefactor\the\spacefactor}\/\fi
  #1\@sf\vfootnote{#1}}
\def\vfootnote#1{\insert\footins\bgroup\eightpoint
  \interlinepenalty\interfootnotelinepenalty
  \splittopskip\ht\strutbox 
  \splitmaxdepth\dp\strutbox \floatingpenalty\@MM
  \leftskip\z@skip \rightskip\z@skip \spaceskip\z@skip \xspaceskip\z@skip
  \textindent{#1}\footstrut\futurelet\next\fo@t}
\def\fo@t{\ifcat\bgroup\noexpand\next \let\next\f@@t
  \else\let\next\f@t\fi \next}
\def\f@@t{\bgroup\aftergroup\@foot\let\next}
\def\f@t#1{#1\@foot}
\def\@foot{\strut\egroup}
\def\footstrut{\vbox to\splittopskip{}}
\skip\footins=\bigskipamount 
\count\footins=1000 
\dimen\footins=8in 
\catcode`@=12 
\newcount\footcount \footcount=0
\def\footn#1{\global\advance\footcount by 1
\footnote{${}^{\number\footcount}$}{#1}}
\def\HB {\hfill\break}
\def\HALF{{\textstyle{1\over 2}}}
\def\AA{{\script A}}
\def\BB{{\script B}}

\def\FF{{\script F}}

\def\JJ{{\script J}}
\def\KK{{\script K}}
\def\LL{{\script L}}

\def\OO{{\script O}}

\def\SS{{\script S}}

\def\WW{{\script W}}

\def\ZZ{{\script ZZ}}
\let\kappa=\varkappa
\let\epsilon=\varepsilon
\let\phi=\varphi 
\let\theta=\vartheta 

\def\QED{\hfill\vbox{\hrule height 0.6pt
	\hbox{\vrule width 0.6pt height 1.8ex \kern 1.8ex
		\vrule width 0.6pt}
	\hrule height 0.6pt}}
\def\real{{\bf R}}

\def\integer{{\bf Z}}

\def\PROOF{\medskip\noindent{\bf Proof.\ }}
\def\REMARK{\medskip\noindent{\bf Remark.\ }}
\def\LIKEREMARK#1{\medskip\noindent{\bf #1.\ }}
\normalbaselineskip=5.25mm
\baselineskip=5.25mm
\parskip=10pt
\beforesectionskipamount=24pt plus8pt minus8pt
\sectionskipamount=3pt plus1pt minus1pt
\def\em{\it}
\tenpoint
\null
\normalbaselineskip=12pt
\baselineskip=12pt
\parskip=0pt
\parindent=22.222pt
\beforesectionskipamount=24pt plus0pt minus6pt
\sectionskipamount=7pt plus3pt minus0pt
\overfullrule=0pt
\hfuzz=2pt
\nopagenumbers
\headline={\ifnum\pageno>1 {\hss\tenrm-\ \folio\ -\hss} \else {\hfill}\fi}
\font\titlefont=ptmb at 14 pt

\font\toplinefont=cmcsc10
\font\pagenumberfont=ptmb at 10pt
\newdimen\itemindent\itemindent=1.5em

\def\textindent#1{\indent\llap{#1\enspace}\ignorespaces}
\def\item{\par\noindent
\hangindent\itemindent\hangafter=1\relax
\setitemmark}
\def\setitemindent#1{\setbox0=\hbox{\ignorespaces#1\unskip\enspace}%
\itemindent=\wd0\relax
\message{|\string\setitemindent: Mark width modified to hold
         |`\string#1' plus an \string\enspace\space gap. }%
}
\def\setitemmark#1{\checkitemmark{#1}%
\hbox to\itemindent{\hss#1\enspace}\ignorespaces}
\def\checkitemmark#1{\setbox0=\hbox{\enspace#1}%
\ifdim\wd0>\itemindent
   \message{|\string\item: Your mark `\string#1' is too wide. }%
\fi}
\setitemindent{3.)}
\expandafter\xdef\csname
sintro\endcsname{1}
\expandafter\xdef\csname
emove\endcsname{(1.1)}
\expandafter\xdef\csname
smain\endcsname{2}
\expandafter\xdef\csname
egl\endcsname{(2.1)}
\expandafter\xdef\csname
ekink\endcsname{(2.2)}
\expandafter\xdef\csname
cinitial\endcsname{Definition\ 2.1}
\expandafter\xdef\csname
fig/NODE/mykonos/users/rougemon/papers/connect/figs/large2.ps\endcsname{Fig.\ 1}
\expandafter\xdef\csname
corthogonal\endcsname{Lemma\ 2.2}
\expandafter\xdef\csname
cmain-thm\endcsname{Theorem\ 2.3}
\expandafter\xdef\csname
sold-paper\endcsname{3}
\expandafter\xdef\csname
ekink-N\endcsname{(3.1)}
\expandafter\xdef\csname
etubeq\endcsname{(3.2)}
\expandafter\xdef\csname
cnorm-z\endcsname{Lemma\ 3.1}
\expandafter\xdef\csname
ezz\endcsname{(3.3)}
\expandafter\xdef\csname
eaa\endcsname{(3.4)}
\expandafter\xdef\csname
cz-set\endcsname{Theorem\ 3.2}
\expandafter\xdef\csname
ccollapse\endcsname{Theorem\ 3.3}
\expandafter\xdef\csname
sapplications\endcsname{4}
\expandafter\xdef\csname
cn-admissible\endcsname{Definition\ 4.1}
\expandafter\xdef\csname
cn-case\endcsname{Theorem\ 4.2}
\expandafter\xdef\csname
cinA\endcsname{Lemma\ 4.3}
\expandafter\xdef\csname
enice-eq\endcsname{(4.1)}
\expandafter\xdef\csname
cnice\endcsname{Definition\ 4.4}
\expandafter\xdef\csname
fig/NODE/mykonos/users/rougemon/papers/connect/figs/tree.ps\endcsname{Fig.\ 2}
\expandafter\xdef\csname
fig/NODE/mykonos/users/rougemon/papers/connect/figs/movie.ps\endcsname{Fig.\ 3}
\expandafter\xdef\csname
eisolate-eq\endcsname{(4.2)}
\expandafter\xdef\csname
cisolate\endcsname{Definition\ 4.5}
\expandafter\xdef\csname
ccoarsening\endcsname{Theorem\ 4.6}
\expandafter\xdef\csname
sproofs\endcsname{5}
\expandafter\xdef\csname
erhs-gl\endcsname{(5.1)}
\expandafter\xdef\csname
eperturbation\endcsname{(5.2)}
\expandafter\xdef\csname
epolynomN\endcsname{(5.3)}
\expandafter\xdef\csname
cold-lemma\endcsname{Lemma\ 5.1}
\expandafter\xdef\csname
cspeed\endcsname{Proposition\ 5.2}
\expandafter\xdef\csname
cinterior\endcsname{Lemma\ 5.3}
\expandafter\xdef\csname
eheat\endcsname{(5.4)}
\expandafter\xdef\csname
cexterior\endcsname{Lemma\ 5.4}
\expandafter\xdef\csname
eDDD1\endcsname{(5.5)}
\expandafter\xdef\csname
el2contraction\endcsname{(5.6)}
\expandafter\xdef\csname
elinfcontr\endcsname{(5.7)}
\expandafter\xdef\csname
ev00\endcsname{(5.8)}
\expandafter\xdef\csname
eapriori\endcsname{(5.9)}
\expandafter\xdef\csname
cresolvent\endcsname{Lemma\ A.1}
\expandafter\xdef\csname
eneumann\endcsname{(A.1)}
\expandafter\xdef\csname
clocal\endcsname{Lemma\ B.1}
\expandafter\xdef\csname
eduhamel\endcsname{(B.1)}
\expandafter\xdef\csname
elin\endcsname{(B.2)}
\expandafter\xdef\csname
el2\endcsname{(B.3)}
\def\3HALF{\textstyle{3\over2}}

\def\norm{\vert\kern-0.1em\vert\kern-0.1em\vert}

\def\min{{\rm min}}
\def\max{{\rm max}}
\def\lpi{{\rm L}^{\!p}(\real,dx)}
\def\ltwo{{\rm L}^{\!2}(\real,dx)}
\def\linf{{\rm L}^{\!\infty}(\real,dx)}
\def\tub{{\cal T}_{N,\Gamma,\sigma }}
\def\ONG{\Omega_{N,\Gamma}}
\def\ZZ{{\script Z}_{N,\Gamma}}
\def\AA{{\script A}_{N,\Gamma}}
\def\ZZZ{{\script Z}_{N-2,\Gamma}}
\def\AAA{{\script A}_{N-2,\Gamma}}
\headline={\ifnum\pageno>1 {\toplinefont Dynamics of embedded kinks}
\hfill{\pagenumberfont\folio}\fi}
{\titlefont{\noindent Dynamics of kinks in the Ginzburg-Landau equation: 
Approach to a metastable shape and collapse of embedded pairs of kinks}}
\vskip 0.5truecm
{\it\centerline{J. Rougemont}}
\vskip 0.3truecm
{\eightpoint\centerline{D\'epartement de Physique Th\'eorique,
Universit\'e de Gen\`eve, CH-1211 Gen\`eve 4, Switzerland}}
\bigskip
\centerline{\eightpoint\bf Abstract}
{\narrower\smallskip\noindent
{\eightpoint\baselineskip 11pt
\noindent We consider initial data for the real Ginzburg-Landau
equation having two widely separated zeros. We 
require these initial conditions to be locally close to a stationary
solution (the ``kink'' solution) except for a perturbation supported 
in a small interval between the two kinks. 
We show that such a perturbation vanishes on a time scale
much shorter than the time scale for the motion of the
kinks. The consequences of this bound, in the context of earlier
studies of the dynamics of kinks in the
Ginzburg-Landau equation, [ER], are as follows: we consider 
initial conditions $v_0$ whose restriction to a bounded interval $I$
have several zeros, not too regularly spaced, and other zeros of $v_0$
are very far from $I$. We show that all these zeros eventually 
disappear by colliding with each other. This relaxation process 
is very slow: it takes a time of order exponential of the length of
$I$.}\smallskip}
\bigskip
\tenrm

\SECT(intro)Introduction 

This paper is a continuation of [ER], where a
model of interface dynamics was analyzed. This model is based on the
Ginzburg-Landau equation in an unbounded one-dimensional domain. A
similar model had originally been studied on a 
finite interval subject to Neumann boundary conditions by J. Carr and
R.L. Pego, [CP1,CP2]. For a physical motivation and a discussion of
related models, see Bray, [B], and references therein.
The interfaces are defined as the zeros of a solution $v(x,t)$ of the 
real Ginzburg-Landau evolution equation. These zeros are shown to have the
following behavior: let their positions on the real line be denoted by
$z_k(t)$, with $z_j(t)<z_{j+1}(t)$, 
$j=0,\dots,N-1$. When the zeros are sufficiently far from each other, 
their dynamics is approximately described by:
$$
\partial_tz_j(t)\,\approx\,E\left
(e^{-\alpha_c(z_{j+1}(t)-z_j(t))}-e^{-\alpha_c(z_j(t)-z_{j-1}(t))}\right
)~,
\EQ(move)
$$
with $E,\alpha_c$ some numerical constants. After some
time, two zeros might come close to each other. Then they annihilate
over a short time scale. The shape of the
function $v(x,t)$ is shown to be essentially determined by the
locations of the zeros, assuming the initial condition $v(x,0)$ to have
``the right shape''. In particular, the interface (hereafter called
``kink'') corresponding to
the zero at $x=z_k$ is very close to the function $\tanh\bigl
(\pm(x-z_k)/\sqrt{2}\bigr )$. For more general
equations than the Ginzburg-Landau equation, similar results hold, but
$E$, $\alpha_c$, and the local shape of the kinks are different.

In the present paper, we discuss the following problem 
left open in [ER]: suppose $v(x,0)$ has four zeros,
$z_1,\dots,z_4$. Suppose that at some time $t=t_1<\infty $, 
$z_2$ and $z_3$ annihilate, by the mechanism explained above. 
Then $v(x,t_1)$ looks as follows (see
\fig(/NODE/mykonos/users/rougemon/papers/connect/figs/large2.ps)):
it has two zeros $z_1(t_1)$ and $z_4(t_1)$, it is (say) positive
in-between and has a ``bump'' in the middle, where $z_2$ and $z_3$
have just annihilated. Does the evolution 
bring this system back to the situation where
$v(x,t)$ has the ``right'' shape for Eq.\equ(move) to hold? Namely,
does the ``bump'' vanish sufficiently fast, so that one sees again two
slowly moving kinks, which might again be shown to
annihilate after some time? We will show below that this is indeed 
the case. This is different from the case of a dynamics in a
bounded spatial domain, since in [CP2], the authors were only able to
show that if one starts with $N$ kinks, then after the collapse of a
pair of them, the number of kinks will never be more than $N-2$, but
they were unable to iterate this result.

\LIKEREMARK{Acknowledgments}This work was supported by the Fonds
National Suisse. It is a pleasure to thank J.-P. Eckmann for useful 
discussions and encouragements.

\SECT(main)Definitions and main result

Our results can easily be extended to any equation of the form
discussed in [ER]. Here, however, we restrict ourselves to the
following real Ginzburg-Landau equation which is the most explicit
example:
$$
\eqalign{
\partial_tv(x,t)\,&=\,\partial_x^2v(x,t)+v(x,t)-v^3(x,t)~,\cr
x\in\real~,&\quad t\in\real^+~,\quad v(x,t)\in[-1,1]~.
}
\EQ(gl)
$$
This equation has
a few simple time-independent solutions which will be used in this
paper: the trivial ones $v(x,t)=\pm 1$, the ``kinks'' $v(x,t)=
\tanh(\pm x/\sqrt{2})$,
and a one-parameter family of periodic solutions $v(x,t)=\phi _D(x)$, 
where $D\in(\pi,\infty )$ is half the period of $\phi_D$ (see [ER],
Proposition 1.1). We fix the definition
of $\phi _D$ by requiring that $\phi _D(x)<0$ for $0<x<D$. Note that
translates of a solution are also solutions of the equation. Since
$v(x,t)=\pm 1$ are solutions of Eq.\equ(gl), 
the maximum principle ([CE], Theorem 25.1) implies that if
$|v(x,0)|\le 1$, then $|v(x,t)|\le 1$ for all $t>0$. Hence the
evolution Eq.\equ(gl) is well-defined.

Throughout the paper, we will 
use the following notations: $\|\cdot\|_p$ is the usual norm of 
$\lpi$, where $dx$ is Lebesgue measure. The scalar product of $\ltwo$
is denoted by $(\cdot,\cdot)$.
If $A\subset\real$ is a Borel set, $\chi_A$
denotes its (sharp) characteristic function and $\Theta_A$ is a
smooth version of it, {\it i.e.,} $\Theta_A(x)=1$ for $x\in
A$, $\Theta_A(x)=0$ if ${\rm dist}(x,A)>1$, and
$\sum_{j=0}^k\|\partial^{j}_x\Theta_A\|_\infty\le C$ for some constant
$C$ independent of $A$ and for a sufficiently large integer $k$. 

Let $Z=\{z_1,z_2\}\in\real^2$, we define $|Z|\equiv z_2-z_1$ and 
$m_1(Z)\equiv (z_1+z_2)/2$. We will always assume $|Z|>\pi$. 
With any such $Z\in\real^2$ we associate a bounded smooth function
$u_Z$ as in [CP1,ER]:
$$
u_Z(x)\,=\,\Theta_{\rm L}(x)\tanh\left ({x-z_1\over\sqrt{2}}\right )
+\Theta_{\rm C}(x)\phi_{|Z|}(x-z_2)
+\Theta_{\rm R}(x)\tanh\left ({z_2-x\over\sqrt{2}}\right )~,
\EQ(kink)
$$
where ${\rm L}=(-\infty,z_1-1/2]$, ${\rm C}=[z_1+1/2,z_2-1/2]$, 
and ${\rm R}=[z_2+1/2,\infty )$.

We next introduce a class of functions containing the initial
conditions we are interested in, see
\fig(/NODE/mykonos/users/rougemon/papers/connect/figs/large2.ps):
\CLAIM Definition(initial) We say that $f:\real\to\real$ is an
$(\epsilon,\alpha,\ell,\Gamma)$--admissible function if $\|f\|_\infty\le1$
and there is a $Z=\{z_1,z_2\}$ in $\real^2$ such that $f$
can be written as $f=u_Z+w_1+w_2$ with:
\item{--}The two kinks are far apart: 
$$
z_2-z_1\,\equiv\,|Z|\,>\,2\Gamma\,>\,2\ell\,>\,\pi~.
$$
\item{--}The large part of the perturbation has support in a
relatively small interval, far from the kinks:
$$
{\rm supp}(w_2)\,\subset\,\left [m_1(Z)-\ell,m_1(Z)+\ell\right]
\equiv Y(\ell)~.
$$
\item{--}The remainder of the perturbation is very small:
$$
\max\left\{\|w_1\|_2,\|w_1\|_\infty\right\}\,\le\,e^{-\alpha |Z|}~.
$$
\item{--}The function is positive between the two kinks: 
$$
f(x)\,>\,\epsilon\quad{\rm ~for~all~}x\in Y(\ell)~.
$$

\figurewithtexplus
/NODE/mykonos/users/rougemon/papers/connect/figs/large2.ps
/NODE/mykonos/users/rougemon/papers/connect/figs/large2.tex 1.7 14.5
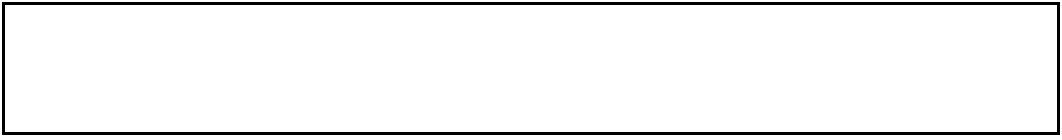
0.0 An admissible function $f$ (full line), with $u_Z$ superimposed
(dotted line).\cr

The next lemma states that 
admissible functions have the following property: 
one can associate with them a
function $u_Z$ as given by Eq.\equ(kink) in such a way that the
difference is ``almost'' in a stable subspace of the linearized
evolution, see \clm(old-lemma) below. Let 
$$
\tau_1(Z,x)\,=\,-\Theta_{(-\infty ,m_1-1]}(x)\partial_xu_Z(x)~,\quad
\tau_2(Z,x)\,=\,-\Theta_{[m_1+1,\infty )}(x)\partial_xu_Z(x)~.
$$
\CLAIM Lemma(orthogonal) For any positive $\alpha,\epsilon,\ell$, 
for sufficiently large $\Gamma<\infty $, if $f$
is an $(\epsilon,\alpha,\ell,\Gamma)$--admissible function, then 
there is a unique $Z'\in\real^2$ with $\bigl
(f-u_{Z'},\tau_j(Z',\cdot)\bigr )=0$, $j=1,2$. Moreover, $Z'$
is a ${\cal C}^2$ function of $f$.

\PROOF Let $\FF(u,Z)\in\real^2$, 
$\bigl (\FF(u,Z)\bigr )_j=\bigl (u-u_Z,\tau_j(Z,\cdot)\bigr )$,
$j=1,2$. Then $\FF(u_Z,Z)=0$ and $D_Z\FF(u_Z,Z)$ is invertible, see
[ER], Lemma 5.3. 
Let $\BB(u_Z,\sigma)$ denote the ball of radius $\sigma$ around
$u_Z$ in the topology
$\norm f\norm_Z=\int|f||\partial_xu_Z|$. Then,
by the Implicit Function Theorem,  for
sufficiently small $\sigma$, there is a ${\cal C}^2$ function
$Z':\BB(u_Z,\sigma)\to\real^2$ such
that $\FF\bigl (u,Z'(u)\bigr )=0$ for all $u\in\BB(u_Z,\sigma)$. Note
that there is a $\Gamma$ such that any
$(\epsilon,\alpha,\ell,\Gamma)$--admissible function $f$ is in this
ball of radius $\sigma$.\QED
\REMARK We will use the following shorthands, to keep the
notation simple: we will always write $Z(t)$ for
$Z'\bigl (v(\cdot,t)\bigr )$. Similarly $m_1(Z)$, defined above as 
$m_1(Z)=(z_1+z_2)/2$, is now a
function of $t$ also, denoted simply by $m_1(t)$.
Throughout the paper, the same letter $C$ will denote
several numerical constants. We will often write the time
variable as a subscript, {\it e.g.,} $v_t(\cdot)\equiv v(\cdot,t)$.

We next state the main technical result of the paper:
\CLAIM Theorem(main-thm) There are constants 
$\alpha_c >0$ and $K,M<\infty $ such that for any positive 
$\epsilon<1$, $\ell<\infty $, $\alpha\le\alpha _c$, 
for sufficiently large $\Gamma=\Gamma
(\epsilon,\ell)<\infty $, if $v_0(x)$ is an 
$(\epsilon,\alpha,\ell,\Gamma)$--admissible function and 
$v_t(x)=v(x,t)$ is the
corresponding solution of Eq.\equ(gl), then there is a $T<K|Z(0)|$ 
for which 
\item{1)}$|Z(T)|>|Z(0)|/2>\Gamma$,
\item{2)}$\max\left\{\|v_T-u_{Z(T)}\|_2,\|v_T-u_{Z(T)}\|_\infty\right\}
\le Me^{-\alpha_c|Z(T)|}$.

\PROOF See \sec(proofs).
\LIKEREMARK{Remark 1}The constant $\alpha _c$ is the same as in
Eq.\equ(move) and is, for the equation considered in this paper, equal to
$\sqrt{2}$. We use the constant $\alpha\le\alpha _c$ in the proofs
because we like to bound $C\exp(-\alpha _c|Z|)$ by $\exp(-\alpha |Z|)$
when some constant $C$ appears.
\LIKEREMARK{Remark 2}The reader must view this result in the following
context: we suppose that at some time $t_0<0$ in the past, $v_{t_0}$ had
four zeros $z_0,\dots,z_3$. Under the  evolution Eq.\equ(gl), these
zeros have moved, until $z_1$ and $z_2$ (the central pair) 
annihilate. We suppose this happens at time $t=0$, {\it i.e.,} 
$u_t$ becomes strictly positive in the interval $(z_0,z_3)$ when 
$t=0$. Such a $u_0$ is the typical admissible function to which we
want to apply \clm(main-thm). \clm(main-thm) imply that after a time
$T$ which is small compared to the time $T'$ needed for a kink to move a
large distance (typically a distance $\Gamma$ needs a time
$T'=\OO(\Gamma\exp(\alpha _c|Z|))\gg T=\OO(|Z|)$), 
the distance (in the topologies of $\linf$ {\it and}
of $\ltwo$) between the solution $v_T$ of
Eq.\equ(gl) and a two-kink state $u_Z$ for some $Z\in\real^2$ will be
smaller than any prefixed 
constant, provided $|Z(0)|$ is large enough. This
shows that the local shape of $u_t$ is restored by the evolution.

\SECT(old-paper)Dynamics of many kinks

In \sec(main), we have restricted ourselves to the case of two
kinks. In \sec(applications), we will extend \clm(main-thm) to the
case of $N+1$ kinks and work out some applications of this result. To
do so we first generalize the definitions of \sec(intro) and recall
some results proved in [ER]. 

Let $N$ be an odd integer  (the case of even $N$ needs only minor 
modifications), let $\ONG$ be the set of all sequences of 
$N+1$ kinks separated by a distance at least $\Gamma$: 
$$
\ONG\,=\,\Bigl\{Z=\{z_0,\dots,z_N\}\in\real^{N+1}\,
:\,z_j-z_{j-1}>\Gamma~,\,j=1,\dots,N\Bigr\}~.
$$
Let $\Gamma>\pi$, $Z\in\ONG$, $z_{-1}=-\infty $, and
$z_{N+1}=+\infty $. We define the following numbers and intervals:
$$
\matrix{
\ell_j\,&=\,z_j-z_{j-1}~,\hfill&j\,=\,0,\dots,N+1~,\cr
|Z|\,&=\,\min\{\ell_1,\dots,\ell_N\}~,\hfill& \cr
m_j\,&=\,\HALF(z_j+z_{j-1})~,\hfill&j\,=\,0,\dots,N+1~,\cr
I_j\,&=\,\bigl (z_{j-1}+\HALF,z_j-\HALF\bigr
)~,\hfill&j\,=\,0,\dots,N+1~.\cr
}
$$
We next construct the analogue of $u_Z(x)$, Eq.\equ(kink), for
the case of $N+1$ kinks:
$$
\eqalign{
u_Z(x)\,&=\,\Theta_{I_0}(x)\tanh\left({x-z_0\over \sqrt{2}}\right )+
\Theta_{I_{N+1}}(x)\tanh\left({z_N-x\over \sqrt{2}}\right )\cr
\,&\,\qquad\qquad+\sum_{j=1}^N
(-1)^j\Theta_{I_j}(x)\phi_{\ell_j}(x-z_{j-1})~.
}
\EQ(kink-N)
$$
The following properties are readily verified: 
$u_Z\in{\cal C}^\infty(\real)$, 
$\partial_x^2u_Z(x)+u_Z(x)-u_Z^3(x)=0$ for $|x-z_j|>1/2$, 
$u_Z(z_j)=0$ for $j=0,\dots,N$, and $(-1)^j\chi_{I_j}(x)u_Z(x)<0$.

Below, we will extend the notion of admissible functions, 
\clm(initial), to the case of $N+1$ kinks. We first define a smaller
set $\tub$ of nice functions, depending on the two parameters 
$\Gamma>\pi$ and $\sigma>0$:
$$
\eqalign{
\tub\,&=\,\Bigl\{v\in L^\infty(\real) : \|v\|_\infty \le 1,\cr
\,&\,\qquad\max\bigl\{\inf\limits_{Z\in\ONG}
\|\partial_x(v-u_Z)\|_2
~,~\inf\limits_{Z\in\ONG}\|v-u_Z\|_2\bigr\}<\sigma
\Bigr\}~.
}
\EQ(tubeq)
$$

We finally introduce a set of $N+1$ functions, each of which
``generates'' the translation of one kink:
$$
\tau_{j}(Z,x)\,=\,-\Theta_{M_j}(x)\partial_xu_Z(x)~,\qquad j=0,\dots,N~,
$$
where $M_j=[m_j+1,m_{j+1}-1]$.

In order to state the main results of [ER], we need to formulate a
lemma, which summarizes several steps of the proofs presented in [ER].
We define $L_Zf\equiv\partial_x^2f+\bigl(1-3u_Z^2\bigr)f$ 
(this is the r.h.s.\ of Eq.\equ(gl) linearized around $v=u_Z$).
\CLAIM Lemma(norm-z) For any integer $N<\infty$, for sufficiently
large $\Gamma$ and sufficiently
small $\sigma$, there exists a unique $~{\cal C}^2$ function 
$Z:\tub\to\ONG$
such that $\bigl (v-u_{Z(v)},\tau_j(Z(v),\cdot)\bigr )=0$ for
$j=0,\dots,N$.
Moreover, there is a constant $M>1$ 
such that for any $v\in\tub$, one has:
$$
M\|L_{Z(v)}w\|_2^2\,\ge\,-\bigl (w,L_{Z(v)}w\bigr )\,\ge\,{1\over M}
\|w\|_2^2~,
$$
where $w=v-u_{Z(v)}$.

The first part of \clm(norm-z) is the analogue of \clm(orthogonal),
with virtually the same proof. The second part is based on the
spectral properties of the self-adjoint operator $L_Z$.
It seems now legitimate to introduce the notation 
$-\bigl (f,L_Zf\bigr)\equiv\|f\|_Z^2$ for $f$ in the orthogonal
complement of ${\rm span}\{\tau_j(Z,\cdot),j=0,\dots,N\}$ in $\ltwo$.

It has been proved in [ER] that there exists a strictly positive function
$g(Z)$, satisfying $g(Z)\to 0$ when $|Z|\to\infty $ such that the
following holds:  
\CLAIM Theorem(z-set) Let 
$$
\ZZ=\bigl\{v\in\tub:\|w(v)\|_{Z(v)}<g\bigl (Z(v)\bigr )\bigr\}~.
\EQ(zz)
$$
For any $N<\infty $, for
sufficiently large $\Gamma$ and sufficiently  small $\sigma$, if
$v_0\in\ZZ$, then either 
the orbit $v_t$ of $v_0$ under Eq.\equ(gl) lies in
$\ZZ$ for all times $t>0$, or
there is a time $T<\infty $ and a
$k\in\{1,\dots,N\}$ such that $\ell_k(v_T)=\Gamma$.\HB 
Moreover there is an $\alpha_c>0$ such that Eq.\equ(move)
holds for $Z(t)=Z(v_t)$ with $v_t\in\ZZ$, in the sense that the
r.h.s.~minus the l.h.s.~is $\OO(e^{-3\alpha_c|Z|/2})$,
and there is an $s>0$ such that the set 
$$
\AA=\bigl\{v\in\tub:\|w(v)\|_{Z(v)}<s\bigr\}
\EQ(aa)
$$ 
is exponentially attracted towards $\ZZ$.

It has also been shown that the above results can be extended
to the case of infinitely many zeros, provided there are numbers $k,N$
such that the intervals
$[z_k,z_{k+1}]$ and $[z_{k+N},z_{k+N+1}]$ are very large compared to
$|Z^*|$, where $Z^*=\{z_{k+j}\}_{j=1,\dots,N}$ (see also
\sec(applications) below). 

Consider an orbit $v_t$ of Eq.\equ(gl) satisfying
$v_t\in\ZZ$ for $t<T<\infty $ and $\ell_2(v_T)=\Gamma$, {\it
i.e.,} the second case of the alternative 
of \clm(z-set) holds with $k=2$. Then,
the following result was proved in [ER]:
\CLAIM Theorem(collapse) For sufficiently large $\Gamma$, there are a
$\Gamma_0>\Gamma$ and a $T_0>T$ such that if $\min\left\{
\ell_1(v_T),\ell_3(v_T)\right\}>\Gamma_0$, then $|v_{T_0}(x)|>0$ for
$x\in[z_0+\Gamma_0/2,z_2-\Gamma_0/2]$.

\SECT(applications)Applications of \clm(main-thm)

The function $v_{T_0}$ of \clm(collapse) is {\it not} in
$\AAA$. This was the main unsatisfactory point with the results of [ER]. 
In this section, we show that after a finite 
time and under some conditions on the position of the remaining kinks, 
it will get into $\AAA$. 

First we state a condition which permits a generalization of
\clm(main-thm) to the case of $N+1$ kinks using the set $\ZZ$ defined
in Eq.\equ(zz).

\CLAIM Definition(n-admissible) Let $f\in\linf$, $\|f\|_\infty \le
1$. We call $f$ admissible if there is a $g\in\ZZ$, an $\ell<\infty $,
and a $j\in\{1,\dots,N\}$ such that $f$ can
be written as $f=g+w$ where:
\item{1)}$w$ has support in $\left[m_j\bigl (Z(g)\bigr )-\ell,m_j\bigl
(Z(g)\bigr )+\ell\right]\equiv Y_j(\ell)$, 
\item{2)}$|f(x)|>\epsilon $ for $x\in Y_j(\ell)$,
\item{3)}there is a $\beta>1$ such that $\beta\ell<|Z|$.

\REMARK Assumption 3) above is only 
stated for future reference. It is just
a different formulation of the statement that for fixed $\ell$, $|Z|$
must be larger than some $\Gamma=\Gamma(\ell)$, see \clm(main-thm).

\CLAIM Theorem(n-case) Let $N<\infty $, let $v_0$ satisfy
\clm(n-admissible). If $\,\Gamma>\pi$ and $\beta>1$ 
are sufficiently large, then the conclusions
of \clm(main-thm) hold for the corresponding solutions $v_t$ of
Eq.\equ(gl) (with $|Z|$ as defined in \sec(old-paper)).

\PROOF The proof of \clm(n-case) is easily worked out by combining
\clm(main-thm), \clm(local), and a maximum principle as in
Eq.\equ(v00) and Eq.\equ(apriori) in the proof of \clm(main-thm). 
The details are left to the reader.\QED

\CLAIM Lemma(inA) Let $v_T$ and $\alpha_c$,
be as in \clm(main-thm). There is 
a $C<\infty $ such that $\|v_T-u_{Z(T)}\|_{Z(T)}\le C\exp\bigl
(-\alpha_c|Z(T)|\bigr )$, where $\|f\|_Z^2=-\bigl(f,L_Zf\bigr)
=-\bigl(f,f''+(1-3u_Z^2)f\bigr)$.

\PROOF See \sec(proofs).

\LIKEREMARK{Remark 1}\clm(inA) shows that the orbit of $v_T$
enters the attracting neighborhood 
$\AAA$ of the invariant set $\ZZZ$, after the collapse of 
an interval (see Eq.\equ(zz) and Eq.\equ(aa) 
for the definitions of these sets). 
In a way, it shows that the basin of attraction of the invariant cone
$\ZZZ$ is much larger than $\AAA$, and in fact contains points that have
just come out of $\ZZ$ through the collapse of a pair of kinks. This
is to be compared with the case of an evolution equation in a 
bounded spatial domain, see Proposition 4.3 in
[CP2]. There it was shown that any orbit reaching the boundary 
of $\ZZ$ cannot ever re-enter it. This only shows that one will never see
again a configuration with $N$ kinks. But one still expects to
see configurations with less kinks. With the result
\clm(inA), we are able to show that there are initial configurations
which ``cascade'' from $\ZZ$ to $\ZZZ$ to ${\script Z}_{N-4,\Gamma}$
and so on.
\LIKEREMARK{Remark 2}In [CP1,CP2], the authors 
use three small parameters in their proofs, and the game with these
three parameters is quite involved. The first one, $\rho$ in their
notations, corresponds to $1/\Gamma$ with our definitions. 
This small parameter is the main ingredient of the whole proof. 
Their second small parameter is the diffusion constant $\epsilon
$ (this {\it is not} the $\epsilon$ of \clm(initial)). Upon rescaling,
this small parameter can be identified with the inverse of the size of
the spatial domain in which the evolution is defined. It can be 
eliminated by
working directly on the infinite line as was shown in [ER]. Finally
the present paper shows that the constraint on the third 
parameter, called $\sigma$ in [CP1], can be relaxed. This parameter 
measures the size of the allowed perturbations around the multi-kink
state $u_Z$ (this {\it is} the $\sigma$ of Eq.\equ(tubeq)).

We next introduce a set of configurations of zeros which are quite
general and for which we can control the dynamics of the kinks for
arbitrarily long times. 
We begin with a construction involving only finitely many zeros,
{\it i.e.,} we give ourselves a $Z\in\ONG$, with $N=2M+1$.
We use the same notations as in \sec(old-paper): $Z=\{z_0,\dots,z_N\}$, 
$\ell_j=z_j-z_{j-1}$, $j=1,\dots,N$, $|Z|=\min\{\ell_1,\dots,\ell_N\}$. 
We will construct a discrete dynamics which approximate the behavior
of the zeros of a solution of Eq.\equ(gl) by using only the following
simple rule: at each time-step, erase the two nearest zeros and keep
the other ones fixed (this {\it is} the model studied by Bray,
Derrida, and Godr\`eche in [BDG]). 
Then we will state conditions on the initial
configuration of zeros which imply that the continuous dynamics of
Eq.\equ(gl) remain well-approximated by this discrete model for a long
enough time.
 
We associate a labeled tree with the configuration $Z$. 
By a labeled tree we mean a set of vertices
and edges. Each vertex is associated (``labeled'') with a number
or with $\infty $. 
The vertices are drawn on $M+1$ levels numbered $0,\dots,M$. 
On level $k$ there are $N-2k+2$ vertices numbered
$0,\dots,N-2k+1$. Hence the $(j+1)^{\rm th}$
vertex from the left on the $(k+1)^{\rm th}$ level from the top is 
identified with $(k,j)\in\integer^2$. 
It is labeled with $v(k,j)\in\real\cup\{\infty \}$ 
which will be defined below. 
There are edges between some vertices of level $k$ and some vertices of
level $k+1$ which will also be constructed below by
iteration. We first define
$$
\matrix{
v(k,0)\,=\,v(k,N-2k+1)\,&=\,\infty ~,\hfill &k\,=\,0,\dots,M~,\cr
\hfill v(0,j)\,&=\,\ell_j~,\hfill &j\,=\,1,\dots,N~.
}
$$ 
We next construct level $k+1$ 
from level $k$, $0\le k\le M-1$. We define $j_\min(k)$ by
$$
v\left(k,j_\min(k)\right)\,=\,\min\{v(k,j):j=1,\dots,N-2k\}~.
$$ 
We suppose here that there is a unique such $j_\min(k)$.
(In \clm(nice) below we
will restrict ourselves to configurations for which this is true.) The
edges are drawn according to the following rule:
\item{1)}There are three edges going from the vertices 
$(k,j_\min(k))$, $(k,j_\min(k)+1)$,
$(k,j_\min(k)-1)$ to the single vertex $(k+1,j_\min(k)-1)$. 
\item{2)}There is an edge between $(k,j)$ and $(k+1,j)$ (if
$j<j_\min(k)-1$) or between $(k,j)$ and $(k+1,j-2)$ (if $j>j_\min(k)+1$).

\noindent It remains to define the numbers $v(k+1,j)$:
$$
v(k+1,j)\,=\,\sum_{m:(k,m)\rightarrow (k+1,j)}v(k,m)~,
$$
where $a\rightarrow b$ means ``there is an edge between $a$ and $b$.''
If one element of the above sum is $\infty $, then the sum is $\infty $.
This construction is iterated from $k=0$ to $k=M-1$. 
We also define sequences $Z^{(k)}=\{z^{(k)}_0,\dots,z^{(k)}_{N-2k}\}$
of real numbers: 
for $k=0$ we simply let $Z^{(0)}=Z$. For $k>0$ we first define
$z_0^{(k)}$ by the following procedure:  
starting from the vertex $(k,1)$ one goes up the
tree following always the leftmost possible edge, until one reaches
vertex $(0,j_0)$. More precisely 
$$
j_0\,=\,\min\bigl\{j\in\{1,\dots,N\}:(k,1){\rm
~is~connected~to~}(0,j){\rm ~by~edges}\bigr\}~.
$$
We let $z^{(k)}_0=z_{j_0-1}^{(0)}$ and 
$z^{(k)}_{j+1}=z^{(k)}_j+v(k,j+1)$, $j=0,\dots,N-2k-1$. 
\CLAIM Definition(nice) We say that $Z\in\ONG$ is 
$(N,\gamma_1)$--non-degenerate if the corresponding tree has labels
$v(k,j)$ which satisfy: for each $k=0,\dots,M-1$, let
$$
d_1(k)\,=\,v(k,j_\min(k)) \quad
d_2(k)\,=\,\min\bigl\{v(k,j):j\in\{1,\dots,N-2k\}\backslash j_\min(k)
\bigr\}~,
$$ 
then 
$$
\gamma_1 d_1(k)\,<\,d_2(k)~,
\EQ(nice-eq)
$$
with $\gamma_1>1$.

The discrete dynamics is now easy to formulate: $Z^{(k)}$ is the
configuration of zeros after $k$ steps of the discrete-time dynamics.
The dynamics ends when there are only two zeros left, namely after $M$
steps. This discrete dynamics is a good approximation for the continuous
dynamics of Eq.\equ(gl) in the sense that if one starts with an
initial condition $v_0$ with a set of zeros equal to $Z^{(0)}$, 
then there are times $t_1<t_2<\dots<t_n$ such that $v(\cdot,t_k)$ 
has zeros approximately given by the set $Z^{(k)}$. In terms of the
continuous dynamics \clm(nice) means that two successive collapse
times are never too close.

\LIKEREMARK{Example}We take $N=7$, $Z=\{0,7,27,32,33,41,44,56\}$ (in
units in which $\Gamma=1$).
\fig(/NODE/mykonos/users/rougemon/papers/connect/figs/tree.ps) shows
the corresponding tree. We obtain the following sets of zeros:
$$
Z^{(1)}\,=\,\{0,7,27,41,44,56\}~,\quad
Z^{(2)}\,=\,\{0,7,27,56\}~,\quad
Z^{(3)}\,=\,\{27,56\}~.
$$
\fig(/NODE/mykonos/users/rougemon/papers/connect/figs/movie.ps) shows the 
zeros $z_j^{(k)}$ on the  interval 
$[-1,57]$ and the functions $u_Z$ given by Eq.\equ(kink-N) with 
$Z=Z^{(k)}$. The numbers $d_j(k)$ are in this case given by:
$$
\matrix{
&d_1(1)\,=\,1~,\hfill &d_1(2)\,=\,3~,\hfill &d_1(3)\,=\,7~,\hfill\cr
&d_2(1)\,=\,3~,\hfill &d_2(2)\,=\,7~,\hfill &d_2(3)\,=\,20~.\hfill\cr
}
$$
For each $k=0,\dots,3$, we have $d_2(k)>2 d_1(k)$. 
Hence $Z$ is $(7,2)$--non-degenerate in the sense of \clm(nice).

\figurewithtexplus
/NODE/mykonos/users/rougemon/papers/connect/figs/tree.ps
/NODE/mykonos/users/rougemon/papers/connect/figs/tree.tex 5.0 8.0
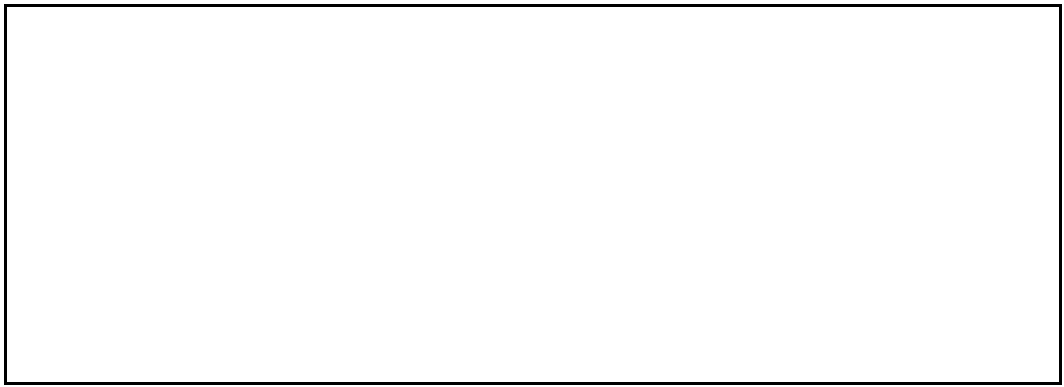
0.0 The tree associated
with the configuration $Z=\{0,7,27,32,33,41,44,56\}$. The numbers are
the labels of the vertices, {\it i.e.,} the distances between two
successive zeros of the functions $u_{Z^{(k)}}$ shown in
\fig(/NODE/mykonos/users/rougemon/papers/connect/figs/movie.ps).\cr
\figurewithtexplus
/NODE/mykonos/users/rougemon/papers/connect/figs/movie.ps
/NODE/mykonos/users/rougemon/papers/connect/figs/movie.tex 6.5 17.0
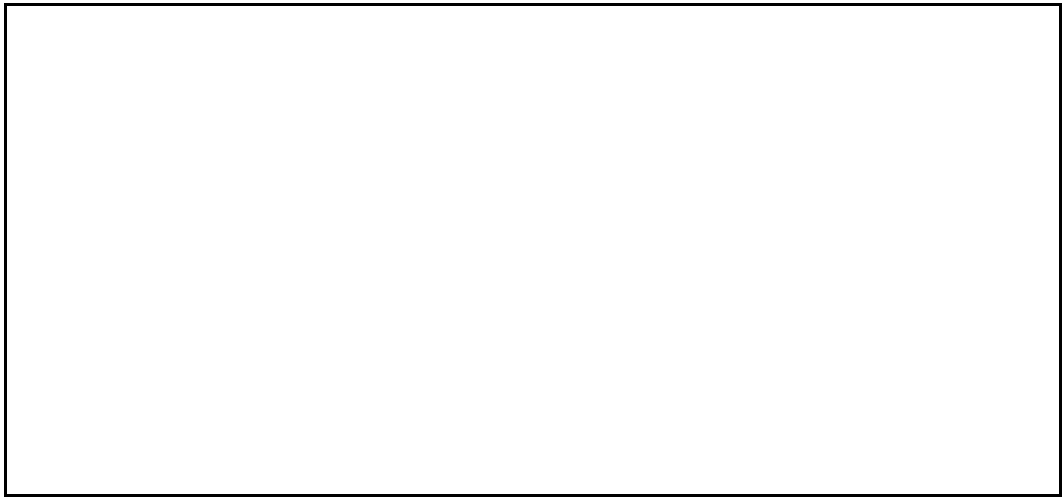
0.0 The horizontal
lines are 4 copies of the interval $[-1,57]$. The vertical lines show
the points $z_j^{(k)}$, $k=0,\dots,3$, $j=0,\dots,7-2k$, with
$j$ going from the left to the 
right and $k$ from top to bottom. The
superimposed dotted lines are the functions $u_{Z^{(k)}}$.\cr

As will be shown in \clm(coarsening), for each $N<\infty $, for
sufficiently large 
$\gamma_1=\gamma_1(N)$, \clm(nice) implies that the discrete
model is a good approximation up to the time when all kinks have
collapsed. Unfortunately, $\gamma_1(N)\to\infty $ when $N\to\infty$. 
Hence \clm(nice) is not a sufficient condition to control
infinitely many kinks. However, since all kinks will disappear in a
finite time we can still make a condition on the remaining (infinitely
many) kinks, so that nothing ``invades'' the small region we are
looking at during this time. This is done by the following definition.
\CLAIM Definition(isolate) We call
$Z_\infty=\{z_j\}_{j\in\integer}\in\real^\integer$, a 
$(k,N,\gamma_1,\gamma_2)$--separable
configuration of zeros if $\{z_k,\dots,z_{k+N}\}$ is in $\ONG$, is
$(N,\gamma_1)$--non-degenerate, and the following holds
$$
\log\left (\min\left\{\ell_k,\ell_{k+N+1}\right \}\right )
\,\ge\,\gamma_2(z_{k+N}-z_k)~,
\EQ(isolate-eq)
$$
with $\gamma_2>1$.

We next describe the ``generic'' behavior of zeros inside a finite
region of the real line.
\CLAIM Theorem(coarsening) Let $N<\infty $.  Let $\Gamma, \ZZ$ be as
in \clm(z-set). For sufficiently large
$\gamma_2=\gamma_2(N)$, $\gamma_1=\gamma_1(N)$, 
for any $(k,N,\gamma_1,\gamma_2)$--separable configuration $Z_\infty $,
the following holds:\HB
Let $\KK=[z_k-\Gamma,z_{k+N}+\Gamma]$ and let $v_t(x)$ be a solution
of Eq.\equ(gl) for which
{\item{1)}$\bigl\{x\in\real:v_0(x)=0\bigr\}=Z_\infty $,
\item{2)}There exists a $v_0^*\in\ZZ$
for which $\bigl (v_0-v_0^*\bigr )\chi_\KK\equiv 0$, with $\ZZ$ as in
\clm(z-set).}\HB
Then there is a $T<\infty
$ such that $|v_T(x)|>0$ for all $x\in\KK$.

\PROOF Let $v_t$ and $v_t^*$ be the orbits of $v_0$ and $v_0^*$ under
Eq.\equ(gl). We suppose $k=0$. 
For any $\delta >0$, and for all $T<\exp\bigl
(\gamma_2(z_N-z_0)\bigr )-\log\bigl (1/\delta\bigr )$, 
for sufficiently large $\gamma_2=\gamma_2(N,\delta)$, 
\clm(local) and \clm(isolate) imply 
$\|\chi_\KK\bigl (v_T-v^*_T\bigr )\|_\infty\le\delta $. Hence it
suffices to prove the claim with $v_t$ replaced by $v_t^*$ and check
that there is a $\gamma_2(N,\delta )$ such that $T$ 
satisfies the above inequality with a sufficiently small $\delta$.

By \clm(nice), Eq.\equ(move), and \clm(collapse), there is a time
$T_1<C\exp\bigl (\alpha_c(z_N-z_0)\bigr )$ such that
$|v_{T_1}^*(x)|>0$ if
$x\in X_j\equiv[z_{j-1}-\Gamma,z_j+\Gamma]$ where $j=j_\min(0)$.
Let $2\epsilon=\inf_{x\in X_j}|v_{T_1}^*(x)|$, let $\beta =\gamma_1$. Then
$\pm v_{T_1}^*$ satisfies \clm(n-admissible) hence by \clm(n-case) and
\clm(inA) there is a time $T_2\le K|Z|\le K(z_N-z_0)$ such that
$v_{T_1+T_2}^*\in\AAA$. 

This argument can be repeated until all zeros of 
$v_t^*$ have disappeared. The $k^{\rm th}$ step takes a time 
$T_1(k)+T_2(k)$, where $T_j(k)$ are as above, namely $T_1(k)$ 
is the time for the interval $\ell_{j_\min(k)}$ to collapse, 
and $T_2(k)$ is the time for $v_t^*$ to enter ${\script
A}_{N-2k,\Gamma}$ after this
collapse. Note that the following bound holds because of \clm(nice):
$$
\left |Z\bigl (T_1(k)+T_2(k)\bigr )\right |-d_1(k)
\,\le\,\sum_{n=1}^kn\gamma_1^{-n}\le{N\over \gamma_1-1}~.
$$ 
Assuming $\gamma_1(N)>1+4N/\Gamma$ make the discrete model still valid
up to this time.

By the above argument, for each $k$, $T_1(k)+T_2(k)\le
C\exp\bigl (\alpha_c(z_N-z_0)\bigr )+K(z_N-z_0)$. This gives a total time 
$T_{\rm tot}=\sum_k T_1(k)+T_2(k)\le 2N\exp\bigl
(2\alpha_c(z_N-z_0)\bigr )$. Assuming 
$\gamma_2(N,\delta )>N^{-1}\Gamma^{-1}\bigl (\log\log(1/\delta
)+\log(2N)\bigr )+2\alpha _c$ we have $T_{\rm tot}\le
\exp\bigl (\gamma_2(z_N-z_0)\bigr )-\log(1/\delta)$. Taking for
example $\delta =1/4$ completes the proof of \clm(coarsening).\QED   

\REMARK A separable configuration $Z$ may contain many
disjoint intervals $\KK_n$, $n\in\JJ$ (with $\JJ$ finite or infinite
denumerable), each one satisfying \clm(nice) with a
sufficiently large $\gamma_1$. In this case, \clm(coarsening) holds
with $\KK$ replaced by any of the $\KK_n$, $n\in\JJ$. 
When $\JJ$ is finite, there is an open set
$\WW\subset\linf$ such that any orbit $v_t$ of Eq.\equ(gl) 
with $v_0\in\WW$ satisfies the conclusions of \clm(coarsening) with
$\KK$ replaced by any $\KK_n$, $n\in\JJ$.

\SECT(proofs)Proofs of \clm(main-thm) and of \clm(inA)

In this section, we consider the case $N=1$ ({\it i.e.,} two kinks) 
as in \sec(intro). The general case is similar, see
\sec(applications). In addition, we denote $\LL(f)$ the r.h.s.\ of
Eq.\equ(gl):
$$
\LL\bigl (f\bigr )(x)\,=\,\partial_x^2f(x)+f(x)-f^3(x)~,
\EQ(rhs-gl)
$$
and $w_t$ is the perturbation of the
pair of kinks, namely, $w_t(x)=v_t(x)-u_{Z(t)}(x)$. One has the equation:
$$
\eqalign{
\partial_tw_t(x)\,&=\,\LL\bigl (u_{Z(t)}\bigr
)(x)-\sum_{i=1,2}\partial_tz_i(t)\partial_{z_i}u_{Z(t)}(x)\cr
\,&\,\quad+\bigl (L_{Z(t)}w_t\bigr )(x)-3u_{Z(t)}(x)w_t^2(x)-w_t^3(x)~,
}
\EQ(perturbation)
$$
where 
$$
\bigl (L_Zf\bigr )(x)\,=\,\partial_x^2f(x)+\bigl (1-3u_Z^2(x)\bigr )f(x)~.
$$
The above differential expression defines a self-adjoint operator with
domain $D(L_Z)$ dense in $\ltwo$. The same
symbol, $L_Z$, will be used for this operator.

We will also use the notation $N(f,g)$ for the following polynomial
appearing on the r.h.s.\ of Eq.\equ(perturbation):
$$
N(f,g)\,=\,1-3f^2-3fg-g^2~.\EQ(polynomN)
$$

The following results are taken from [ER].
\CLAIM Lemma(old-lemma) There are constants $\alpha_c>0$, $M^*>0$,
$C<\infty $ such that for sufficiently
large $|Z|$, the following holds:
\item{1)}$\|\LL(u_Z)\|_\infty\le Ce^{-\alpha_c |Z|}$.
\item{2)}$\|L_Z\tau_k(Z,\cdot)\|_2\le Ce^{-\alpha_c|Z|/2}$, for $k=1,2$.
\item{3)}For $k=1,2$, if $|x-z_k|\le |Z|/2$, then
$$
\left |u_Z(x)-\tanh\Bigl({(-1)^k(z_k-x)\over\sqrt{2}}\Bigr)\right |
\,\le\,Ce^{-\alpha_c |Z|}~.
$$
\item{4)}If $w\in D(L_Z)$ satisfies $(w,\tau_k(Z,\cdot)\bigr)=0$, for
$k=1,2$, then 
$$
\bigl (w,L_Zw\bigr )\,\le\,-M^*\bigl(w,w\bigr)~.
$$
\item{5)}For $k=1,2$, $M_k={\rm supp}(\tau_k(Z,\cdot))$, 
$$
\|\chi_{M_k}\bigl (\partial_{z_k}u_Z-\tau_k(Z,\cdot)\bigr
)\|_\infty \,\le\,Ce^{-\alpha _c|Z|}~.
$$

\REMARK Statement 4) above is a direct consequence of the
spectral analysis of $L_Z$ performed in [ER] (see also 
\clm(norm-z), set $1/M\approx M^*$). In more
intuitive words, statements 1) and 2) say that $u_Z$ is almost a
stationary solution of Eq.\equ(gl), statement 3) shows that the
interface (locally) looks like the kink solution and statement 4)
shows that
the perturbation $w$, when defined as in \clm(orthogonal) is
(nearly) orthogonal to the unstable manifold of the point $|Z|=\infty
$. Statement 5) shows that $\tau_j$ is close to the generator of the
translation of the $j^{\rm th}$ kink. 

The strategy of the proof of \clm(main-thm)
is the following: first we show that solutions $v_t$ of
Eq.\equ(gl) that {\it remain} admissible for all $t\in[0,T]$ must
satisfy: the speed of the kinks
$\partial_tz_i(t)$ is very small, the ``large part'' of the
perturbation ($w_2$ in the notation of \clm(initial)) decays uniformly
and that the ``small part'' ($w_1$) remains small. 
Then we use the maximum principle and 
an inductive argument to show that admissible initial conditions
remain admissible {\it and} converge to a small ball around $u_{Z(t)}$.

We begin with a bound on the speed of the kinks.
\CLAIM Proposition(speed) Let $v_t$ be a solution of Eq.\equ(gl). 
For any $\alpha\le\alpha_c$, if  $v_{t^*}$ is
$(\epsilon,\alpha,\ell,\Gamma)$--admissible for some 
$t^*>0$, $\ell<\infty $, $\epsilon >0$, and
sufficiently large $\Gamma=\Gamma(\alpha)$, then $z_i(t^*)$ satisfies:
$$
|\partial_tz_i(t^*)|\,\le\,Ce^{-\alpha|Z(t^*)|}~.
$$

\PROOF For simplicity, we write $t$ for $t^*$. We also write
$\tau_j(t)$ for the function $\tau_j(Z(t),\cdot)$. By the definition of
$Z(t)\equiv Z(v_t)$, see \clm(orthogonal), we have
$\partial_t(w_t,\tau_j(t))=0$, or
$$
\sum_{i=1,2}\partial_tz_i(t)\left\{
\bigl(\partial_{z_i}u_{Z(t)},\tau_j(t)\bigr)-
\bigl(w_t,\partial_{z_i}\tau_j(t)\bigr )
\right \}
\,=\,\bigl (\partial_tv_t,\tau_j(t)\bigr )~.
$$
If we define
$\SS_{ij}=\bigl(\partial_{z_i}u_{Z(t)},\tau_j(t)\bigr)
-\bigl(w_t,\partial_{z_i}\tau_j(t)\bigr)$ then the matrix 
$\SS=\bigl (\SS_{ij}\bigr)_{i,j=1,2}$ is invertible with uniformly
bounded inverse (see [ER]). Thus we can write
$$
\eqalign{
|\partial_tz_i(t)|\,&=\,\left |\sum_{j=1,2}\SS^{-1}_{ij}\bigl
(\partial_tv_t,\tau_j(t)\bigr )\right |\cr
\,&=\,\left |\sum_{j=1,2}\SS^{-1}_{ij}\bigl
(\LL(u_{Z(t)})+L_{Z(t)}w_t-3u_{Z(t)}w_t^2-w_t^2,
\tau_j(t)\bigr )\right |\cr
\,&\le\,C\biggl(\|\LL(u_{Z(t)})\|_2\sup_j\|\tau_j(t)\|_2
+\sup_j\left |\bigl (w_t,L_{Z(t)}\tau_j(t)\bigr )\right |\cr
\,&\phantom{\le\,C\biggl(}+\|(1-\chi_\KK)
w_t\|_2^2+\sup_j\|\chi_\KK\,\tau_j(t)\|_\infty\biggr)~,
}
$$
where $\chi_\KK$ is the characteristic function of the interval
$\KK=[m_1(t)-\ell,m_1(t)+\ell]$, and $\LL(\cdot)$ is given by
Eq.\equ(rhs-gl). Using \clm(old-lemma), one finds that
each term in the above expression is bounded by $C\exp\bigl
(-\alpha|Z(t)|\bigr )$.\QED

We next prove two lemmas which establish bounds on the evolution in
the middle of the interval enclosed by the two kinks (first lemma) and
outside and near the boundary of this interval (second lemma).
\CLAIM Lemma(interior) For any $\alpha\le\alpha _c$, $\epsilon >0$,
$\ell<\infty $, and
sufficiently large $\Gamma=\Gamma(\alpha,\epsilon,\ell)>0$, if $v_t$ is an
$(\epsilon,\alpha,\ell,\Gamma)$--admissible solution of Eq.\equ(gl)
for all $t\le1$, then 
$$
\|\chi_\KK\bigl (u_{Z(T)}-v_T\bigr )\|_\infty \,\le\,C\bigl
(e^{-\alpha|Z(0)|}+\|\chi_\KK (u_{Z(0)}-v_0)\|_\infty e^{-\epsilon T}
+\sup_{0\le t\le T}\|\chi_\Delta w_t\|_\infty \bigr )
$$
for $T\le 1$ and 
for any $\KK=\left[m_1(0)-\ell^*,m_1(0)+\ell^*\right]$,
$\Delta={\rm supp}(\Theta'_\KK)$, where $\ell+1\le\ell^*\le\Gamma/2$.

\PROOF By \clm(speed), we can take $\Gamma$ so large that for all $t\le1$,
$|m_1(t)-m_1(0)|\le 1$ and $||Z(t)|-|Z(0)||\le 1$. 
We use the notation $N(f,g)$ defined in
Eq.\equ(polynomN). For $x\in\KK^*\equiv\KK\cup\Delta={\rm
supp}(\Theta_\KK)$, one has 
$$
\eqalign{
\,&\,N(u_{Z(t)}(x),w_t(x))\cr
\,&=\,1-u_{Z(t)}^2(x)-2u_{Z(t)}(x)\bigl
(u_{Z(t)}(x)+w_t(x)\bigr )
-w_t(x)\bigl (u_{Z(t)}(x)+w_t(x)\bigr )\cr
\,&\le\,1-(1-Ce^{-\alpha\Gamma/2})^2-2\epsilon
(1-Ce^{-\alpha\Gamma/2})-\epsilon \cr
\,&\le\,-\epsilon ~,
}
$$
provided $\Gamma$ is such that $C\exp\bigl (-\alpha\Gamma/2\bigr )
\le\epsilon/2$. We introduce the heat kernel
$$
G_t(x)\,=\,{1\over\sqrt{4\pi t}}\exp\left ({-x^2\over 4t}\right )~.
\EQ(heat)
$$ 
We next use
Eq.\equ(perturbation), \clm(resolvent), and Lemma 7 of [C] to obtain
the following for $x\in\KK$:
$$
\eqalign{
&|w_T(x)|\,=\,|\Theta_{\KK}(x)w_T(x)|\cr
&\,=\,\Bigl|\int_{-\infty
}^\infty\!\!\!dy\,G_T(x-y)w_0(y)\Theta_{\KK}(y)+\int_0^T\!ds
\int_{-\infty}^\infty\!\!\!dy\,G_{T-s}(x-y)\cr
&\,\phantom{=\,}\times\biggl( 
N\bigl (u_{Z(s)}(y),w_s(y)\bigr
)w_s(y)\Theta_{\KK}(y)-\Theta_{\KK}'(y)w_s'(y)
-\Theta_{\KK}''(y)w_s(y)\cr
&\,\phantom{=\,\times\biggl(}+\Theta_{\KK}(y)\LL\bigl
(u_{Z(s)}\bigr)(y)-\sum_{i=1,2}\partial_t
z_i(s)\Theta_{\KK}(y)\partial_{z_i}u_{Z(s)}(y)\biggr)\Bigr|\cr
&\,\le\,\Bigl|\int_{-\infty }^\infty\!\!\!dy\,
G_T(x-y)\Theta_{\KK}(y)w_0(y)
-\epsilon\int_0^T\!ds\int_{-\infty}^\infty
\!\!\!dy\,G_{T-s}(x-y)\Theta_{\KK}(y)w_s(y)\Bigr|\cr
&\,\,\quad +\Bigl|\int_0^T\!ds\int_{-\infty }^\infty 
\!\!\!dy\,G_{T-s}'(x-y)\Theta_{\KK}'(y)w_s(y)\Bigr|
+C\sup_{0\le s\le T}e^{-\alpha |Z(s)|}\cr
&\,\le\,e^{-\epsilon T}\left|\int_{-\infty}^\infty\!\!\!dy\,
G_T(x-y)\Theta_{\KK}(y)w_0(y)\right|
+C\sup_{0<s<T}\|\chi_\Delta w_s\|_\infty
+Ce^{-\alpha |Z(0)|}~.
}
$$\QED

\CLAIM Lemma(exterior) For any $\alpha\le\alpha_c$, 
$\delta >1$, $\epsilon >0$, $\ell<\infty$, and
sufficiently large $\Gamma=\Gamma(\alpha,\ell,\delta)>0$, if $v_0$
is an $(\epsilon,\alpha,\ell,\Gamma)$--admissible function
then the corresponding solution $v_t$ of Eq.\equ(gl) satisfies:
$$
\max\left\{\|\chi_\BB\bigl (u_{Z(t)}-v_t\bigr)\|_2~,
~\|\chi_\BB\bigl (u_{Z(t)}-v_t\bigr )\|_\infty\right\}
 \,\le\,C\sup_{0\le s\le t}\bigl (e^{-\alpha_c|Z(s)|}+\|\chi_\Delta
w_s\|_\infty\bigr )
$$
for any $\BB=\bigl(-\infty ,m_1(0)-\hat\ell\bigr]
\cup\bigl[m_1(0)+\hat\ell,\infty\bigr)$ where 
$\hat\ell\ge\ell+\delta$, $\Delta={\rm
supp}(\Theta'_\BB)$, and $t\le1$.

\PROOF We take $\Gamma$ so large that, using \clm(speed),
$|m_1(t)-m_1(0)|<(\delta -1)$ for all $t\le 1$. We let
$\BB^*=\BB\cup\Delta$.
\LIKEREMARK{Bound on $\|\cdot\|_2$}By Eq.\equ(perturbation), we have
$$
\eqalign{
\HALF\partial_t\|\Theta_\BB w_t\|_2^2\,&=\,
\bigl (\Theta_\BB\LL(u_{Z(t)}),\Theta_\BB w_t\bigr)
-\sum_{i=1,2}\partial_t z_i(t)
\bigl(\Theta_\BB\partial_{z_i}u_{Z(t)},\Theta_\BB w_t\bigr )\cr
\,&\,\quad
+\bigl(\Theta_\BB w_t,\Theta_\BB L_{Z(t)}w_t\bigr )
-\bigl(\Theta_\BB (3u_{Z(t)}+w_t)w_t^2,\Theta_\BB w_t\bigr )\cr
\,&\le\,Ce^{-\alpha_c|Z(t)|}\|\Theta_\BB w_t\|_2
+\bigl (\Theta_\BB w_t,L_{Z(t)}\Theta_\BB w_t\bigr )\cr
\,&\,\quad
+\bigl (\Theta_\BB w_t,-2\Theta'_\BB w_t'-\Theta_\BB''w_t\bigr )
+4\|\chi_{\BB^*}w_t\|_\infty\|\Theta_\BB w_t\|_2^2\cr
\,&\le\,Ce^{-\alpha_c|Z(t)|}\|\Theta_\BB w_t\|_2
+C\|\Theta_\BB w_t\|_2\|\chi_\Delta w_t\|_\infty\cr
\,&\,\quad
-\Bigl\{M^*-4\|\chi_{\BB^*}w_t\|_\infty-\sum_{i=1,2}
\bigl(\partial_{z_i}u_Z,\Theta_\BB w_t\bigr )^2\Bigr\}
\|\Theta_\BB w_t\|_2^2~,
}\EQ(DDD1)
$$
using the spectral properties of the linear operator $L_{Z(t)}$,
\clm(old-lemma). Obviously (see \clm(local)) there is a $K<\infty $
such that $\|\chi_{\BB^*}w_t\|_\infty\le\exp(Kt)\|\chi_{\BB^*}
w_0\|_\infty$. Using \clm(old-lemma) and taking $\Gamma$ so large that
for all $t\le1$, 
$$
\eqalign{
&4e^{Kt}\|\chi_{\BB^*}w_0\|_\infty+\sum_{i=1,2}
\bigl(\partial_{z_i}u_Z,\Theta_\BB w_t\bigr )^2\cr
\,&=\,4e^{Kt}\|\chi_{\BB^*}w_0\|_\infty+\sum_{i=1,2}
\bigl(\tau_i,(\Theta_\BB-1) w_t\bigr )^2+Ce^{-\alpha _c|Z|}\cr
\,&\le\,4e^{K-\alpha\Gamma}+Ce^{-\alpha_c\Gamma}
\,\le\,{M^*\over 2}~,
}
$$ 
we can integrate Eq.\equ(DDD1), and we get for all $t\le1$,
$$
\|\Theta_\BB w_t\|_2\,\le\,
C\sup_{0\le s\le t}\bigl (e^{-\alpha_c|Z(s)|}+\|\chi_\Delta
w_s\|_\infty\bigr )+e^{-M^*t/2}\|\Theta_\BB w_0\|_2~.
\EQ(l2contraction)
$$

\LIKEREMARK{Bound on $\|\cdot\|_\infty $}Let $x$ be in $\BB$, 
let $G_t(\cdot)$ be given by Eq.\equ(heat). We get
$$
\eqalign{
|w_t(x)|\,&\le\,\left |\int_{-\infty }^\infty\!dy\,\Theta_\BB(y)
G_t(x-y)w_0(y)\right |\cr
\,&\,\quad+C\left |\int_0^t\!ds\,\int_{-\infty }^\infty
\!dy\,\bigl
(\Theta_\BB(y)G_{t-s}(x-y)+\Theta_\BB'(y)G_{t-s}'(x-y)
\bigr )w_s(y)\right |\cr
\,&\,\quad+C\sup_{0\le s\le t}e^{-\alpha_c|Z(s)|}\cr
\,&\le\,C\sup_{0\le s\le t}\bigl
(e^{-\alpha_c|Z(s)|}+\|\Theta_\BB w_s\|_2+\|\chi_\Delta w_s\|_2\bigr )~.
}
\EQ(linfcontr)
$$
We apply the bound \equ(l2contraction) and \clm(exterior) is proved.
\QED

\LIKEREMARK{Proof of \clm(main-thm)}Assume $v_0$ is
$(\epsilon,\alpha,\ell,\Gamma)$--admissible, for given $\ell$,
$\epsilon $, $\alpha\le\alpha _c$, and for sufficiently large 
$\Gamma$. Let $Z(v_0)=\{z_1(v_0),z_2(v_0)\}$ be given by
\clm(orthogonal). There is a $Y=\{y_1,y_2,y_3,y_4\}\in\real^4$,
$z_1(v_0)<y_1<y_2<y_3<y_4<z_2(v_0)$, with
the following property: define the intervals ${\rm L}=(-\infty
,y_1-1/2)$, ${\rm C}_j=(y_j+1/2,y_{j+1}-1/2)$, 
and ${\rm R}=(y_4+1/2,\infty )$. Then the function
$$
\eqalign{
v_0^*(x)\,&=\,
\Theta_{\rm L}(x)\tanh\left({x-y_1\over\sqrt{2}}\right)+
\Theta_{\rm R}(x)\tanh\left({y_4-x\over\sqrt{2}}\right)\cr
\,&\,\qquad\qquad+\sum_{j=1}^3(-1)^j
\Theta_{{\rm C}_j}(x)\phi_{y_j-y_{j-1}}(x-y_j)~
}
\EQ(v00)
$$
lies strictly below $v_0$. By the maximum
principle, [CE], the orbits $v_t$ and $v_t^*$ of Eq.\equ(gl)
with initial conditions $v_0$ and $v_0^*$
satisfy $v^*_t(x)<v_t(x)$ for all $(x,t)\in\real\times\real^+$. 
Moreover, for sufficiently large $\Gamma$, $Y$ can be chosen such
that $v_0^*$ satisfies the hypotheses of 
\clm(z-set). It follows that there are positive constants $C,B_1,B_2$
and a function $Z^*(t)=\{z_1^*(t),\dots,z_4^*(t)\}
:[0,t^*]\to\real^4$, where $t^*=\exp\bigl(
B_1|z_3^*(0)-z_2^*(0)|\bigr )$ and $Z^*(0)=Y$ such that:
$$
\eqalign{
|z_j^*(t)-z_j^*(0)|\,&\le\,Ce^{-B_2|z_3^*(0)-z_2^*(0)|}~,j=1,\dots,4~,\cr
\|v_t^*-v_t^{**}\|_\infty\,&\le\,Ce^{-B_2|z_3^*(0)-z_2^*(0)|}~,
}
\EQ(apriori)
$$
with $v_t^{**}$ given by Eq.\equ(v00) replacing $Y$ by $Z^*(t)$.

The above discussion shows that if 
there is an $\ell(t)$ such that $v_t$ is 
$(\epsilon,\alpha,\ell(t),\Gamma)$--admissible for $t\le t^*$, 
then $\ell(t)\le\ell_\max\equiv
|z_3^*(0)-z_2^*(0)|+C\exp(-B_2|z_3^*(0)-z_2^*(0)|)$, which
is independent of $\Gamma$, $\alpha $, and $\epsilon $. We choose 
$\Gamma$ such that $\ell_\max\le\Gamma/2$. Moreover, when
using \clm(exterior) and \clm(interior),  we have the bound 
$\|\chi_\Delta w_t\|_\infty\le C\exp(-\alpha _c|Z(0)|/4)$ for all
times $t\le t^*$.

Since $v_0$ is $(\epsilon,\alpha,\ell,\Gamma)$--admissible, by
continuity, there is a time $t>0$ such that
$v_s(x)>\epsilon /2$ for $|x-m_1(0)|<\ell_\max$ and $s<t$. By
\clm(interior), $v_t(x)>\epsilon$ for $|x-m_1(0)|<\ell_\max$. 

We repeat this argument until $t=T_1\equiv |Z(0)|/\epsilon$. It 
follows that for all $t\le T_1$, $v_t$ is
$(\epsilon,\alpha,\ell_\max,|Z(0)|-\delta)$--admissible for 
some $\delta\le CT_1\exp\bigl (-\alpha|Z(0)|\bigr )\le |Z(0)|/3$ 
(this bound follows from \clm(speed) if $\Gamma\le |Z(0)|$ 
is sufficiently 
large).

Using repeatedly \clm(interior) and \clm(exterior), we obtain
$$
\|u_{Z(T_1)}-v_{T_1}\|_2\,\le\, CT_1\sup_{t\le T_1}
e^{-\alpha_c|Z(0)|/4}\quad{\rm and}\quad\|u_{Z(t)}-v_t\|_\infty\,\le\,
CT_1\sup_{t\le T_1}e^{-\alpha_c|Z(0)|/4}~.
$$
We finally use \clm(exterior) with $\BB=\real$ (in fact equations
\equ(l2contraction) and \equ(linfcontr), with $\chi_\Delta\equiv 0$
and $\Theta_\BB\equiv 1$), to show that after a time $T_2$ of order
$|Z(T_1)|$, we get the bound we claimed. The bound on $T=T_1+T_2$
follows from \clm(speed).\QED

We finish this section with the
\LIKEREMARK{Proof of \clm(inA)}We have trivially
$\|f\|_Z^2\le\|f'\|_2^2+4\|f\|_2^2$, hence we need only bound
$\|w_T'\|_2^2$. We decompose
$w_T(x)=\chi_{\rm E}(x)w_T(x)+\chi_{\rm I}(x)w_T(x)$ where
${\rm I}=[m_1(0)-\ell_\max,m_1(0)+\ell_\max]$ and 
${\rm E}=\real\backslash{\rm I}$, with $\ell$ as in \clm(main-thm). Let 
$G_t(x)$, $N(\cdot,\cdot)$, $\LL(\cdot)$ be given by Eq.\equ(heat),
Eq.\equ(polynomN), and Eq.\equ(rhs-gl) respectively. 
We compute first
$$
\eqalign{
\|\chi_{\rm E}w_T'\|_2\,&\le\,
\left\|\int_0^T\!ds\,\int_{-\infty}^\infty\!dy\,
G_{T-s}'(x-y)\Theta_{\rm E}(y)\LL\bigl (u_{Z(s)}\bigr )(y)\right \|_2\cr
\,&\phantom{\le\,}+\left\|\int_{\infty }^\infty \!dy\,G_T'(x-y)\Theta_{\rm
E}(y)w_0(y)\right\|_2\cr
\,&\phantom{\le\,}+
\biggl\|\int_0^T\!ds\,\int_{-\infty }^\infty\!dy\,G_{T-s}'(x-y)\cr
\,&\phantom{\le\,+\|}\times\Bigl
(N\bigl (u_{Z(s)}(y),w_s(y)\bigr )\Theta_{\rm E}(y)w_s(y)-\bigl
(\Theta_{\rm E}'(y)w_s(y)\bigr )'\Bigr)\biggr\|_2\cr
\,&\le\,C\sup_{0\le s\le T}e^{-\alpha |Z(s)|}
+C\sup_{0\le s\le T}\|\chi_{\rm E^*}w_s\|_2~,
}
$$
where ${\rm E^*}=\real\backslash[m_1(0)-\ell_\max+1,m_1(0)+\ell_\max-1]$. 
The remaining term, 
$$
\|\chi_{\rm I}w_T'\|_2\,\le\,\sqrt{\ell_\max}\|\chi_{\rm
I}w_T'\|_\infty ~,
$$
can be bounded by a similar argument as in \clm(interior).\QED

\def\actualnumber{A}
\SECTIONNONR Appendix A

\CLAIM Lemma(resolvent) Let $f(x,t)\in{\rm L}^{\!\infty}\left
(t\in[0,1]\to\ltwo\right )$, let $\lambda\in\real$, let $G_t(x)$ be as in
Eq.\equ(heat). Then, the following holds:
$$
\eqalign{
\,&\,\int_{-\infty}^\infty\!\!dy\,G_t(x-y)f(y,0)
+\lambda\int_0^t\!ds\,\int_{-\infty}^\infty\!\!dy\,G_{t-s}(x-y)f(y,s)\cr
\,&=\,e^{\lambda t}\int_{-\infty}^\infty\!\!dy\,G_t(x-y)f(y,0)~.
}
$$

\PROOF Let $A$ be the following operator, densely defined on 
${\rm L}^{\!\infty}\left
(t\in[0,1]\to\ltwo\right )$:
$$
\bigl (Af\bigr )(x,t)\,=\,\int_0^t\!ds\,\int_{-\infty }^\infty
\!\!dy\,G_{t-s}(x-y)f(y,s)~.
$$
It is easy to see that there is a constant $C$ such that $\|A^nf\|\le
C\|f\|/n!$ which implies that the series
$$
f(x,t)\,=\,\sum_{n=0}^\infty \lambda ^n\bigl (A^ng\bigr )(x,t)
\EQ(neumann)
$$
converges and is a solution of the equation
$$
f(x,t)\,=\,g(x,t)+\lambda \bigl (Af\bigr )(x,t)~.
$$
Substituting $g(x,t)=\int\!dy\,G_t(x-y)f(y,0)$ into Eq.\equ(neumann) 
and using
$$
\int_{\real^2}\!dy\,dz\,G_{t-s}(x-y)G_{s-\ell}(y-z)f(z,\ell)\,=\,
\int_\real\!dy\,G_{t-\ell}(x-y)f(y,\ell)~,
$$
one obtains that 
$$
\eqalign{
f(x,t)\,&=\,\sum_{n=0}^\infty \lambda
^n\int_0^t\!dt_1\,\int_0^{t_1}\!dt_2\,\dots\int_0^{t_{n-1}}\!dt_n\,
\int_{-\infty}^\infty \!\!dy\,G_t(x-y)f(y,0)\cr
\,&=\,e^{\lambda t}\int_{-\infty }^\infty \!\!dy\,G_t(x-y)f(y,0)
}
$$
is a solution of the equation 
$$
f(x,t)\,=\,\int_{-\infty}^\infty\!\!dy\,G_t(x-y)f(y,0)
+\lambda\int_0^t\!ds\,\int_{-\infty}^\infty\!\!dy\,G_{t-s}(x-y)f(y,s)~.
$$
\QED
\def\actualnumber{B}
\SECTIONNONR Appendix B

\CLAIM Lemma(local) Let $\KK=[-L,L]$, let
$\epsilon\in(0,1)$, and let $v_t$ and $w_t$ be two solutions of
Eq.\equ(gl), with $\|\chi_\KK\bigl (v_0-w_0\bigr )\|_\infty
\le\epsilon$. There are $K_1,K_2>0$ such that for any
$\delta,t,\ell$ satisfying 
$$
1\,<\,t\,<\,K_1\bigl (\min\{\ell,\log\epsilon ^{-1}\}
-\log\delta ^{-1}\bigr )~,\quad\ell<L~,
$$
one has 
$$
\|\chi_{[-L+\ell,L-\ell]}\bigl (v_t-w_t\bigr )\|_\infty
\,\le\,K_2(\delta+\epsilon)~.
$$

\PROOF Let $G_t(x)$ be given by Eq.\equ(heat). By Duhamel's principle,
with $F(x,y)=1+x^2+y^2+xy$, we have
$$
v_t(x)-w_t(x)\,=\,\bigl (G_t\star (v_0-w_0)\bigr )(x)
+\int_0^t\!ds\,\Bigl(G_{t-s}\star\bigl (F(v_s,w_s)(v_s-w_s)\bigr)
\Bigr)(x)~,
\EQ(duhamel)
$$
where $\star$ denotes convolution. We next consider
$x\in[-L+\ell,L-\ell]$. For $t>1$, the first term of Eq.\equ(duhamel) 
is easily bounded:
$$
\left|\bigl (G_t\star (v_0-w_0)\bigr )(x)\right |
\,\le\,C_1(e^{-C_2\ell^2/t}+\epsilon)~.
\EQ(lin)
$$ 
We introduce the following notations: $\delta v_t(x)=v_t(x)-w_t(x)$, 
$\phi(y)=\exp\left (-2\sqrt{1+y^2}\right )$, $\phi_x(y)=\phi(x-y)$. 
We first compute
$$
\eqalign{
\,&\,\partial_t\int\phi_x(\delta v_t)^2\cr
\,&=\,2\int\phi_x\delta v_t\Bigl(\partial_y^2\bigl(\delta v_t\bigr )+
F\bigl (v_t,w_t\bigr )\delta v_t\Bigr)\cr
\,&\le\,
-\int\phi_x\bigl (\partial_y(\delta v_t)\bigr )^2
+\int\phi_x(\delta v_t)^2\left (2\sup_{|x|\le 1,|y|\le 1}
|F(x,y)|+\|\phi_x'\phi_x^{-1}\|_\infty ^2\right)\cr
\,&\le\,C_3\int\phi_x(\delta v_t)^2~.
}
$$
This shows that 
$$
\left (\int\phi_x\bigl (\delta v_t\bigr )^2\right)^\HALF
\,\le\,
e^{C_3t}\left (\int\phi_x(\delta v_0(y)\bigr )^2\right )^\HALF
\,\le\,C_4e^{C_3t}\bigl (\epsilon+e^{-\ell}\bigr )~.
\EQ(l2)
$$
Combining Eq.\equ(duhamel), Eq.\equ(l2), and Eq.\equ(lin), 
we obtain the following bound:
$$
\eqalign{
&|v_t(x)-w_t(x)|\,\le\,C_1\bigl (e^{-C_2\ell^2/t}+\epsilon\bigr )\cr
&+\int_0^t\!ds\,\int_\real\!dy\,G_{t-s}(x-y)\phi_x^{1/2}(y)
\phi_x^{-1/2}(y)
\bigl|\delta v_s(y)F\bigl (v_s(y),w_s(y)\bigr )\bigr|\cr
&\le\,C_1\bigl (e^{-C_2\ell^2/t}+\epsilon\bigr )\cr
&+C_5\int_0^t\!ds\,\left(\int_\real\!dy\,\phi_x^{-1}(y)
G^2_{t-s}(x-y)\right)^\HALF
\!\left(\int_\real\!dy\,\phi_x(y)(\delta v_s(y))^2\right )^\HALF\cr
&\le\,C_1\bigl (e^{-C_2\ell^2/t}+\epsilon\bigr )
+C_6\int_0^t\!ds\,e^{C_2(t-s)}\bigl (\epsilon+e^{-\ell}\bigr )\cr
&\le\,C_7\bigl (e^{-C_2\ell^2/t}+\epsilon
+e^{C_2t}(\epsilon+e^{-\ell})\bigr )~.
}
$$
The claim follows easily.\QED

\SECTIONNONR References
 
\eightpoint
\raggedright
\widestlabel{[XXX]}
\vskip 15pt

\ref 
  \no B
  \by Bray, A.J.
  \paper Theory of phase-ordering kinetics
  \jour Adv. Phys.
  \vol 43 {\rm (3)}
  \pages 357--459
  \yr 1994
\endref

\ref
\no BDG
  \by Bray, A.J., Derrida, B. and C. Godr\`eche
  \paper Non-trivial Algebraic Decay in a Soluble Model of Coarsening
  \jour Europhys. Lett.
  \vol 27 {\rm (3)}
  \pages 175--180
  \yr 1994
\endref

\ref
  \no C
  \by Collet, P.
  \paper Thermodynamic limit of the Ginzburg-Landau equations 
  \jour Nonlinearity
  \vol 7 {\rm (4)}
  \pages 1175--1190\break
  \yr 1994
\endref

\ref
  \no CE
  \by Collet, P. and J.-P. Eckmann
  \book Instabilities and Fronts in Extended Systems
  \publisher Princeton, Princeton University Press
  \yr 1990
\endref

\ref 
\no CP1
  \by Carr, J. and R.L. Pego
  \paper Metastable Patterns in Solutions of $u_t=\epsilon^2u_{xx}-f(u)$
  \jour Comm. Pure Appl. Math.
  \vol XLII
  \pages 523--576
  \yr 1989
\endref

\ref 
\no CP2
  \by Carr, J. and R.L. Pego
  \paper Invariant Manifolds for Metastable Patterns in
  $u_t=\epsilon^2u_{xx}-f(u)$
  \jour Proc. Roy. Soc. Edinburgh
  \vol 116A
  \pages 133--160
  \yr 1990
\endref

\ref 
\no ER
  \by Eckmann, J.-P. and J. Rougemont
  \paper Coarsening by Ginzburg-Landau dynamics
  \jour Comm. Math. Phys.
  \toappear 
\endref
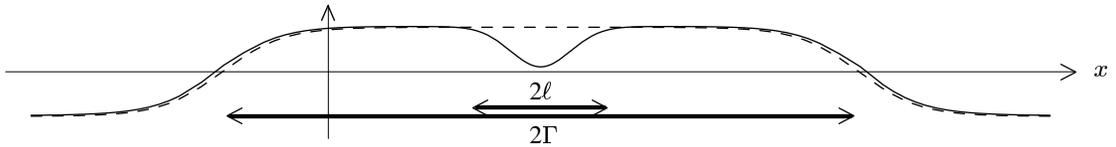
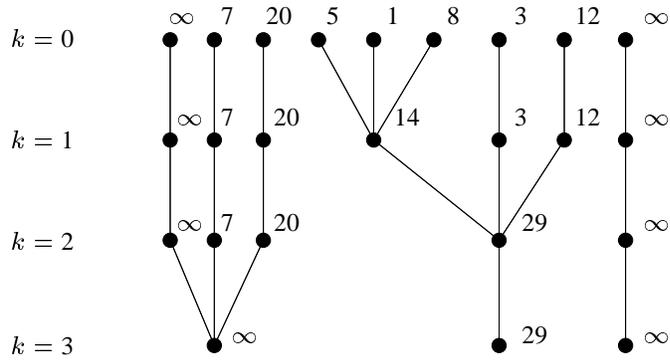
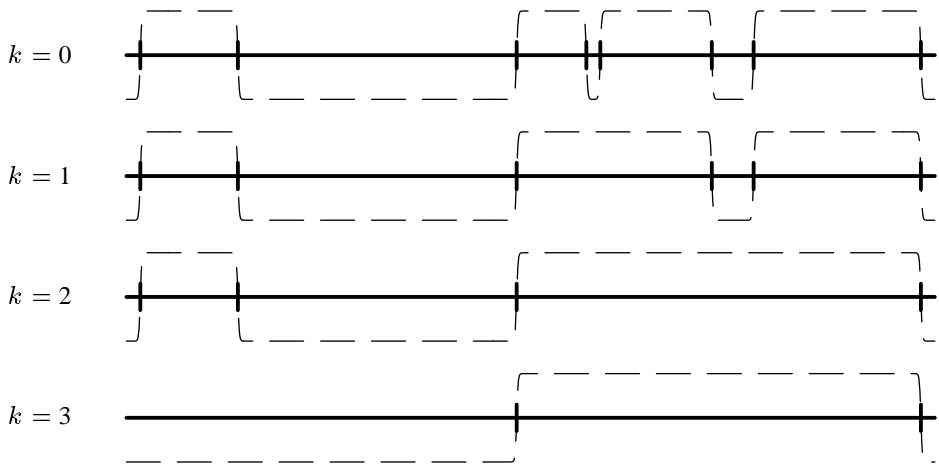
\bye